\documentclass[12pt]{iopconfser}
\usepackage[dvips]{graphics}
\usepackage{graphicx}
\usepackage{bm}
\usepackage{amsmath}
\usepackage{amsfonts}
\usepackage{amssymb}
\usepackage{xcolor}
\usepackage{subfigure}
\usepackage{tabularx}
\usepackage{multirow}
\usepackage{braket}
\usepackage{commath}
\usepackage{color}
\usepackage[colorlinks,linkcolor=blue,anchorcolor=blue,citecolor=blue,urlcolor=blue]
{hyperref}

\begin{document}

\title{Spinless topological chirality from Umklapp scattering in twisted 3D structures}

\author{Cong Chen$^{1,2}$, Xu-Tao Zeng$^{3}$ and Wang Yao$^{1,2\ast}$}

\affil{$^1$New Cornerstone Science Laboratory, Department of Physics, University of Hong Kong, Hong Kong, China}
\affil{$^2$HKU-UCAS Joint Institute of Theoretical and Computational Physics at Hong Kong, Hong Kong, China}
\affil{$^3$School of Physics, Beihang University, Beijing 100191, China}

\email{wangyao@hku.hk}

\begin{abstract}
Spinless systems exhibit unique topological characteristics compared to spinful ones, stemming from their distinct algebra. Without chiral interactions typically linked to spin, an intriguing yet unexplored interplay between topological and structural chirality may be anticipated. Here we discover spinless topological chiralities solely from structural chiralities that lie in the 3D spatial patterning of structureless units, exemplified using two types of twisted graphite systems. In a 3D screw twisted structure without periodicity in all directions, we find a chiral Weyl semimetal phase where bulk topology and chiral surface states are both determined by the screw direction. And in a 3D periodic structure formed with layer-alternating twist angle signs, a higher-order Dirac semimetal with chiral hinge states is discovered. Underlying these novel topological states is the intervalley Umklapp scattering that captures the chirality of the twisted interfaces, leading effectively to a sign-flipped chiral interlayer hopping, thereby introducing $\pi$-flux $\mathbb{Z}_2$ lattice gauge field that alters the symmetry algebra. Our findings point to a new pathway for engineering topological chirality through patterning twisted arrays of featureless units, which can expand the design principles for topological photonics and acoustics.
\end{abstract}

\newpage

\section{Introduction}

Chirality, a fundamental concept across physics, chemistry, and biology~\cite{chiral_chemistry,cahn1966specification,francotte2006chirality}, describes the geometric property of objects that cannot be superimposed onto their mirror images. 
In chemistry and biology, chirality typically pertains to the structures  seen in molecules or proteins that break all the mirror, inversion, or other roto-inversion symmetries. In physics, the concept of chirality also takes into account particles' internal quantum degrees of freedom, such as spin, which transform under spatial operations. Chirality plays a key role in the topological characterization of materials~\cite{Hasan2010RMP,SCZhang2011RMP,chiu2016RMP,Bansil2016RMP,BernivigTopological,ShunQingShen_TI,Ashivin2018RMP}, describing momentum space electronic structures within the crystal bulk as well as on surfaces and edges.
Nontrivial topological chirality often emerges from chiral interactions, such as spin-orbit couplings (SOC)~\cite{HJZhang2009TI,Yazyev2010PRL}. Examples of this include the chiral surface states in topological insulators (TIs)~\cite{Hsieh2009_TI_Science,Souma_PRL_2011,PanZH_PRL_2011}, and the intrinsic chirality of Weyl fermions in topological semimetals~\cite{Weyl_PRB_Wan,WengHM_PRX_2015,Dai2016Weyl,burkov2016TSM,SAY2016Spin,volovik2003universe,YanBH_Weyl_2017}. Additionally, there are instances where the interplay of SOC and structural chirality leads to a correlation between structural and topological chirality~\cite{Chang2018NM,Hassan_2019_Nature,DingHongNC_2019,Science_2020_ChiralTSM,Hasan2021NRM}.

Spinless systems constitute another important context for investigating topological phases of matter, e.g. light element crystals with negligible SOC. Artificial crystals such as photonic and acoustic ones are generically spinless as well, although effective SOC
and topological chirality can be tailored through complex design of elementary units featuring pseudospins~\cite{NatPho_2014_Marin,RMP_2019_Ozawa,BLZhang_2022_NRM}. Spinless systems exhibit distinct topological properties due to their adherence to different symmetry algebra~\cite{YXZhao_2017_PRL,jbYang_2018_PRL,YXZhao_2020_Boundary,YXZhao_2021_PRLSwitch}. For example, spinless systems obey the algebra of time reversal (TR) symmetry $T^2=1$, whereas spinful systems follow $T^2=-1$, leading to different topological classifications~\cite{Classification_PRB_2008,kitaev2009periodic,ZDWang013PRL,YXZhao_2021_PRLSwitch}. 
Moreover, TR symmetric spinless systems  inherently possess $\mathbb{Z}_2$ gauge fields, i.e., the hopping amplitudes being real numbers with either positive or negative values. 
Notably, the $\mathbb{Z}_2$ gauge fields with $\pi$ flux can lead to design of novel topological phases such as 
2D M\"{o}bius insulators~\cite{Mobuis_2020_PRB,Mobuis_2022_PRL,Qiu_acoustic_PRL_2022}, Klein bottle insulators~\cite{Klein_2023_PRB}, higher-order topological semimetals~\cite{HOSM_2021_YXZhao,YXZhao_2021_PRLSwitch}, and mirror Chern insulators~\cite{SpinlessMCI_2023_PRB,wang2024mirror}. 
In reality, $\pi$-flux $\mathbb{Z}_2$ gauge fields and other forms of chiral interactions can be highly nontrivial to realize with only spinless orbitals.  
In the absence of chiral interactions, the manifestation of topological chirality necessitates an alternative cause of chiral symmetry from the spatial patterning of spinless units. This possibility, however, has seldom been explored.

Here we show a new pathway to engineer $\pi$-flux $\mathbb{Z}_2$ gauge field and topological chirality in spinless systems by exploiting {\emph{intervalley} Umklapp scattering in twisted 3D structures.
Using graphite (or 3D graphene) as an example, topological chiralities purely from structural chiralities are demonstrated in two types of twisted patternings. Type-I patterning has adjacent graphene layers all twisted with the same commensurate angle, forming a 3D helical structure lacking translational symmetry in all directions that needs a generalized Bloch theorem to describe. 
It features a unique 3D Weyl semimetal phase, with the bulk topology as well as chiral surface states solely determined by the screw direction. Type-II structure has alternating signs of twist angles for adjacent interfaces and features a higher-order Dirac semimetal phase with chiral hinge states. Underlying these novel topological states is a sign-flipped chiral interlayer hopping, effectively realized by the {\emph{intervalley} Umklapp process that naturally captures the chirality of the interface. Notably, such coupling introduces $\pi$-flux $\mathbb{Z}_2$ lattice gauge field that alters the symmetry algebra, giving rise to the observed topological chirality. 
Our findings unveil a novel approach to achieve varieties of topological chirality-based functionalities through  patterning twisted arrays of featureless units, suggesting new design principles for topological photonics and acoustics.

\section{Results}

The results are organized as follows. We start with the spinless chiral Weyl semimetal phase, by first presenting a description based on a simplified  model with effective chiral interlayer hopping on an untwisted hexagonal lattice. The projective symmetry algebra and the crucial role played by the $\pi$-flux $\mathbb{Z}_2$ gauge field are analyzed.
We then establish the equivalence between the artificial chiral interlayer hopping in the untwisted structure and the realistic \emph{intervalley} Umklapp coupling at commensurately twisted interfaces. This sets the ground for the realization of the spinless chiral Weyl semimetal in a 3D helical structure of the twisted graphite lattice, for which we develop an atomistic Slater-Koster tight-binding calculation based on a generalized Bloch theorem with screw rotational symmetry. Next, as another example of topological chirality from structural chirality, we present the realization of a higher-order Dirac semimetal phase with chiral hinge states in a 3D periodic structure with alternating signs of twist angles for adjacent interfaces.
Lastly, we discuss the robustness of the topological properties against possible forms of disorders.

\subsection{Sign-flipped interlayer hopping and spinless chiral Weyl semimetal}

To break all the in-plane mirror symmetries while preserving in-plane rotational symmetries and time reversal symmetry on an untwisted hexagonal lattice with an infinite number of layers in the $z$-direction (Figure~\ref{TBmodel}(a)), an effective sign-flipped interlayer hopping should be introduced (Figure~\ref{TBmodel}(f)). This chiral interlayer hopping can exhibit two distinct configurations along the $z$ direction, both shown in Figure~\ref{TBmodel}(b) labeled as type-I and type-II.
We will focus on the type-I configuration in this part, and discuss the case of type-II later.
In the Bloch basis of $(\psi_{A}, \psi_{B})^{T}$, the simplified tight-binding (sim-TB) model with the effective chiral interlayer hopping reads,
\begin{equation}
\begin{aligned}
\mathcal{H}_{\text{I}}^{\text{3D}}(\mathbf{k}_{\|},k_z) =& \chi_1 (\mathbf{k}_{\|} ) \sigma_x+\chi_2 (\mathbf{k}_{\|} ) \sigma_y +2 \cos \left(k_z d\right) M  \sigma_0 \\ & +2 \text{i} \sin \left(k_z  d\right) \zeta\lambda (\mathbf{k}_{\|}) \sigma_z,
\end{aligned}
\label{typeA-TB}
\end{equation}
where $\mathbf{k}_{\|}=(k_x, k_y)$, and $\sigma_i$ are Pauli matrices acting on the $A$ and $B$ sublattices. The first line is just the standard 3D $AAA$ graphite model.
Here $\chi_1+\text{i} \chi_2=t_1 \sum_{i=1}^3 \text{e}^{\text{i} \mathbf{k}_{\|} \cdot \mathbf{\delta}_i}$, where $\mathbf{\delta}_1=\frac{1}{3} \mathbf{a}_1+\frac{2}{3} \mathbf{a}_2, \mathbf{\delta}_2=-\frac{2}{3} \mathbf{a}_1-\frac{1}{3} \mathbf{a}_2, \text { and }\mathbf{\delta}_3=\frac{1}{3} \mathbf{a}_1-\frac{1}{3} \mathbf{a}_2$ are the nearest-neighbor intralayer hopping vectors with hopping amplitude $t_1$. 
The second line describes a chiral interlayer hopping, where $\lambda(\mathbf{k}_{\|})=2 \text{i} \lambda_0 \sum_{i=1}^3 \sin (\mathbf{k}_{\|} \cdot \mathbf{d}_i )$ ($\mathbf{d}_1=\mathbf{a}_1$, $\mathbf{d}_2=\mathbf{a}_2$, and $\mathbf{d}_3=-\mathbf{a}_1-\mathbf{a}_2$), and $\zeta=+$ or $-$. 
With $C_{2z}T$ in spinless systems, only real hopping amplitudes are permitted.

\begin{figure}[t]
\setlength{\abovecaptionskip}{0.cm}
\includegraphics[width=10 cm]{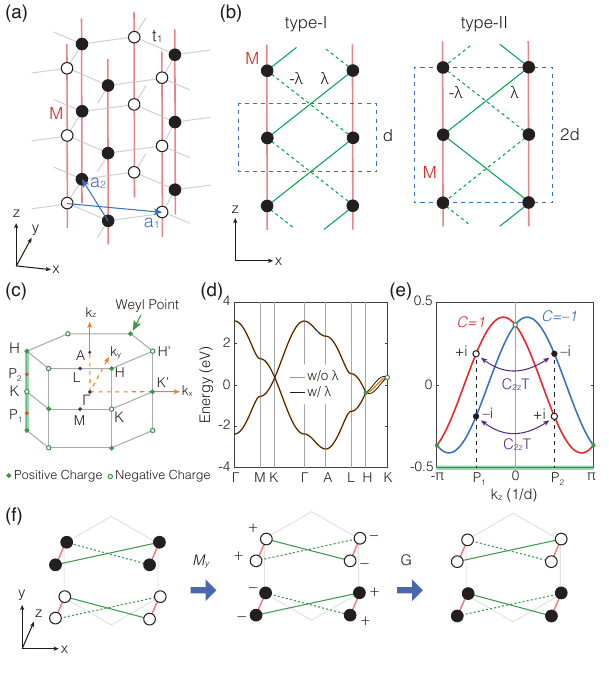}
\centering
\caption{Topolocal chirality from sign-flipped interlayer hopping on an untwisted hexagonal lattice. 
(a) The hexagonal lattice with infinite numbers of layers in the $z$-direction, with solid and hollow dots representing $A$ and $B$ sublattices, respectively. 
(b) Side view of the lattice illustrating the chiral interlayer hopping in two possible configurations, with solid and dashed lines indicating positive and negative hopping amplitudes, respectively.  
(c) Corresponding Brillouin zone. 
(d) Bulk band structures without and with the chiral interlayer hopping $\lambda$ (in type-I configuration). A finite $\lambda$ introduces a splitting along the H-K path. 
(e) Low-energy band structure along the H-K path marked in (c), with the red and blue lines indicating Chern numbers, $C=1$ and $C=-1$, respectively. The labels indicate the eigenvalues of $\mathcal{M}_y$. The $\mathrm{P_1}$ and $\mathrm{P_2}$ points are connected by $C_{2z}T$, exhibiting opposite $\mathcal{M}_y$ eigenvalues. 
(f) Schematic figure showcasing the invariance under the $proper$ mirror symmetry $\mathcal{M}_y=\text{G}M_y$. The signs on the right figures specify the gauge transformation \text{G}. The red and green lines form closed hopping loops that enclose $\pi$-flux.} 
\label{TBmodel}
\end{figure}

Figure~\ref{TBmodel}(d) shows the band structures with and without chiral interlayer hopping. 
Note that the chiral interlayer hopping differs for the $A$ and $B$ sublattices, resulting in the splitting of sublattice degeneracy along the $\mathrm{H}-\mathrm{K}$ paths. 
Thereby leads to the emergence of Weyl nodes located at the corners of the 3D Brillouin zone (BZ) (Figure~\ref{TBmodel}(c)).
We find that these bulk Weyl nodes are topologically nontrivial.  And at each fixed $k_z$, the effective 2D subsystem corresponds to the topological Haldane model~\cite{Haldane_1998}, except for $k_z = 0$ and $k_z = \pm \pi/d$.

To begin, assume $\zeta=+$ for simplicity, for a given $k_z$, a reduced 2D subsystem are denoted by $\mathcal{H}(\mathbf{k}_\|,k_z)$, the interlayer hopping can be described as second nearest neighbor hopping with a complex hopping coefficient of $2\text{i} \sin(k_z d) \lambda_0 $.
When $k_z=0$ or $\pm \pi/d$, the Hamiltonian simplifies to $\mathcal{H}(\mathbf{k}_{\|}, 0)=2 M \sigma_0+\chi_1(\mathbf{k}_{\|}) \sigma_x+\chi_2(\mathbf{k}_{\|}) \sigma_y$ or $\mathcal{H}(\mathbf{k}_{\|}, \pm \pi/d)=-2 M \sigma_0+\chi_1(\mathbf{k}_{\|}) \sigma_x+\chi_2(\mathbf{k}_{\|}) \sigma_y$. These  are just the 2D graphene model with an overall energy shift of $\pm 2M$.
At $k_z=\pi/(2d)$, corresponding to a 2D $k_x-k_y$ plane containing the $\mathrm{P_2}$ point in Figure~\ref{TBmodel}(c), the next nearest neighbor hopping coefficient becomes $2\text{i}\lambda_0$, akin to the magnetic flux in the Haldane model.
Hence it exhibits a nontrivial topological charge of $C=+1$.
We find that any 2D subsystem with $\pi/d >k_z>0$ is topologically nontrivial with $C=+1$.
And any 2D subsystem with $-\pi/d < k_z<0$ features a reversed chiral charge of $C=-1$. This can be verified for $k_z=-\pi/(2d)$, where the next nearest neighbor hopping coefficient becomes $-2\text{i}\lambda_0$.
Therefore, $k_z$ acts as a tuning parameter for the chiral topological phase, and the critical points, namely the $\mathrm{H}$ and $\mathrm{K}$ points, must exhibit band crossing points with opposite chirality.

Initially, we simplify the analysis by considering $\zeta=+$. 
Symmetry analysis reveals that both $\zeta=+$ and $\zeta=-$ are allowed, and interestingly, the band structures are identical for both cases.
Now, let us explore the effects of $\zeta$. 
On one hand, from symmetry perspective, we find: $M_y \mathcal{H}_{I}(\zeta) M_y^{-1}=\mathcal{H}_{I}(-\zeta)$, where $M_y$ represents a vertical mirror reflection perpendicular to the $xz$-plane. 
It implies that reversing the sign of $\zeta$ is equivalent to a spatial mirror reflection.
On the other hand,  $\zeta$ represents the sign of the effective next nearest neighbor hopping coefficient, and thus the sign of the chiral topological charge. 
In other words, flipping the sign of $\zeta$ alters the chirality of all the Weyl points.
This is the unique characteristic of a spinless chiral Weyl semimetal.

\subsection{Projective symmetry algebra of the chiral Weyl semimetal phase}

Breaking spatial in-plane mirror symmetries results in a sign-flipped interlayer hopping, which assigns the lattice gauge field to  certain $\pi$-flux $\mathbb{Z}_2$ gauge field. 
Usually, the symmetries with $\pi$-flux $\mathbb{Z}_2$ gauge field should follow a \emph{projective} algebra, which fundamentally alters the algebraic structure of the symmetry group~\cite{Mobuis_2020_PRB,HOSM_2021_YXZhao,YXZhao_2021_PRLSwitch}.
In the following, we will ascertain the symmetry condition of the underlying chiral Weyl semimetal phase and elucidate the crucial role played by the $\pi$-flux $\mathbb{Z}_2$ gauge field.

First, we focus on symmetry along the $\mathrm{H}$-$\mathrm{K}$ path as shown in Figure~\ref{TBmodel}(c), where the nontrivial band splitting occurs. Along this path, the intralayer terms become zero. Although the model is not invariant under spatial mirror reflection $M_y$, it can be transformed into an equivalent configuration (i.e., another gauge choice) by applying a $\mathbb{Z}_2$ gauge transformation $\text{G}$. 
This transformation involves assigning a sign of $+1$ or $-1$ to each basis at each site. Consequently, the gauge-connection configuration becomes invariant under the so-called \emph{proper} mirror operator, $\mathcal{M}_y=\mathrm{G} M_y$, which is a combination of the gauge transformation and the spatial mirror reflection.
Since both ${M}_y$ and $\text{G}$ are real matrices, it follows that $[\mathcal{M}_y, C_{2z}T]=0$. Moreover, $M_y$ reverses the signs at all sites for $\text{G}$, indicating $\{\mathrm{G}, M_y\}=0$. Additionally, we have $M_y^2=\mathrm{G}^2=1$. Therefore, we can deduce $M_y= \sigma_x$, $\mathrm{G}= \sigma_z$, and $\mathcal{M}_y=\mathrm{G} M_y=i \sigma_y$. This leads to the following algebraic relations:
\begin{equation}
\label{condition}
[C_{2z}T, \mathcal{M}_y]=0, \quad \mathcal{M}_y^2=-1.
\end{equation}
Next, the momentum-space Hamiltonian $\mathcal{H}(k_z)$ along $\mathcal{M}_y$-invariant path, specifically the $\mathrm{H}$-$\mathrm{K}$ path as shown in Figure~\ref{TBmodel}(c), can be represented as a block diagonal form:
\begin{equation}
\label{blockdiag}
\mathcal{H}(k_z)=\left[\begin{array}{cc}
h_{+}(k_z) & 0 \\
0 & h_{-}(k_z)
\end{array}\right],
\end{equation}
where $h_{+}(k_z)$ and $h_{-}(k_z)$ denote the Hamiltonian in the mirror-even ($+\text{i}$) and mirror-odd ($-\text{i}$) subspaces respectively.  
$C_{2z}T$ exchanges the two eigenspaces of $\mathcal{M}_y$. For $\mathcal{M}_y\left|\psi_{ \pm}\right\rangle=\pm \text{i} \left|\psi_{ \pm}\right\rangle$, one can show from the commutation relation and the anti-unitary nature of $C_{2z}T$ that $\mathcal{M}_y C_{2z}T \left|\psi_{ \pm}\right\rangle=\mp \text{i} C_{2z}T \left|\psi_{ \pm}\right\rangle$.
Then, we must have $u h_{+}^*(k_z) u^{\dagger}=h_{-}(-k_z)$, where $u$ is a unitary matrix determined by $C_{2z}T$. In other words, $C_{2z}T$ transforms $\left|\psi_{\pm}, \pm k_z\right\rangle$ into $\left|\psi_{\mp}, \mp k_z\right\rangle$.
For a given $k_z$, the gapped 2D subsystem $\mathcal{H}(\mathbf{k}_{\|},k_z)$ can also be divided into two non-crossing parts, denoted as $\mathcal{H}_{\pm}(\mathbf{k}_{\|},k_z)$, each of which are evolved from $h_{\pm}(k_z)$ on the high symmetry path. The 2D subsystems $\mathcal{H}_+(\mathbf{k}_{\|},k_z)$ and $\mathcal{H}_-(\mathbf{k}_{\|},-k_z)$, which are related by $C_{2z}T$, must possess opposite Chern numbers, since $C_{2z}T$  reverses the Chern number.
Figure.~\ref{TBmodel}(e) illustrates the distribution of $\mathcal{M}_y$ eigenvalues and Chern numbers along $\mathcal{M}_y$-invariant paths. 
Each block $\mathcal{H}_{\pm}$ can exhibit a nontrivial Chern number, and $C_{2z}T$ connects them.

In the above analysis, we see that the exchange of the eigenspace of $\mathcal{M}_y$ by $C_{2z}T$ is crucial for the nontrivial chiral topology. In a scenario where $\mathcal{M}_y^2=+1$, which is typical for most spinless systems without $\pi$-flux $\mathbb{Z}_2$ gauge field, $C_{2z}T$ would preserve the eigenspaces of $\mathcal{M}_y$. This preservation occurs because the eigenvalues $\pm$ of $\mathcal{M}_y$ are real numbers that commute with $C_{2z}T$. Even though we can still write $\mathcal{H}(\mathbf{k})$ in the block diagonal form for eigenspaces of $\pm 1$, the states $\left|\psi_{\pm}, \pm k_z\right\rangle$ are related to $\left|\psi_{\pm}, \mp k_z\right\rangle$ by $C_{2z}T$. As a result, they each must have a zero Chern number.
Breaking all the spatial in-plane mirror symmetries is necessary to fulfill $\mathcal{M}^2=-1$, whereas the \emph{proper} mirror symmetry is restored by introducing a $\mathbb{Z}_2$ gauge transformation.

\subsection{Realization though intervalley Umklapp scattering}
\label{Umklapp_tBG}

The key challenge of realizing such spinless topological phase lies in the coexistence of both positive and negative hoppings. 
While some strategies have been proposed to manipulate the sign of coupling in lattice models~\cite{SignControl_PRL_2016,YXZhao_2021_PRLSwitch}, there remains a dearth of realistic electronic examples that exhibit topological states related to $\pi$-flux $\mathbb{Z}_2$ gauge field.
For the 3D chiral Weyl semimetal model concerned, the role of the negative hopping is to break all the mirror symmetries upon the interlayer hybridization between the massless Dirac cones. We note that the symmetry breaking role can be alternatively played by a twisted interface, which may imprint its structural chirality to the electronic coupling.

There has been extensive literature on small angle twist regime concerning the formation of flat minibands through interlayer hybridization between modestly displayed Dirac cones at the first BZ corners~\cite{moireReviewEvaMacDonaldNatMater2020}, where only \emph{intravalley} channels need to be considered for the interlayer hopping (c.f. Appendix C).
In the scenario of large twist angles, however, the Dirac points from the same valley of adjacent layers get widely separated in momentum space (c.f. Figure~\ref{tBG2D}(a)), so that \emph{intravalley} channel can only hybridize states far away from the Dirac points (where Dirac cones from adjacent layers intersect, c.f. Figure~\ref{tBG2D}(b)). Low-energy sector near the Dirac points is negligibly affected by intravalley channels due to the large energy detuning of states that can be coupled~\cite{MacD_PRB_2010}.
Nonetheless, at $\theta=21.8^{\circ}$, opposite valleys from adjacent layers align in the second BZ~\cite{Umk_Nat_WY_2019}, and consequently the \emph{intervalley} Umklapp channel dominates the layer hybridization in the low-energy sector (c.f. Appendix C).

\begin{figure}[t!]
\includegraphics[width=14 cm]{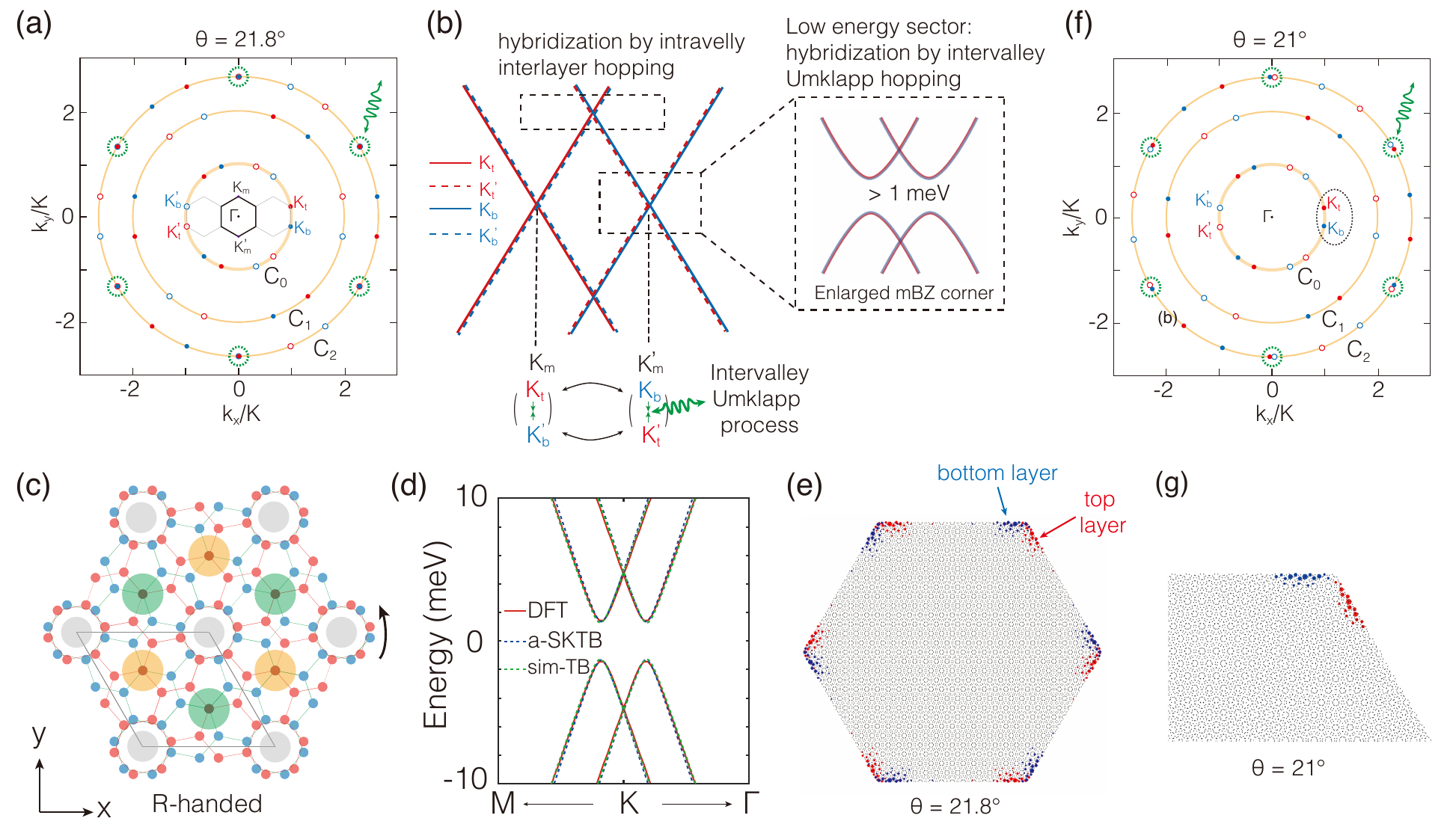}
\centering
\caption{(a) Extended zone scheme at $\theta=21.8^{\circ}$. Solid (hollow) dots denote $\mathbf{K}$ ($\mathbf{K}^{\prime}$) points of the monolayers, and the red (blue) color denotes the top (bottom) layer. $\mathbf{K_m}$ and $\mathbf{K_m^{\prime}}$ denotes the two moir\'{e} Brillion zone (mBZ) corners. 
Top layer $\mathbf{K}_t$ valley and bottom layer $\mathbf{K}_b^{\prime}$ valley get folded to $\mathrm{K_m}$ mBZ corner.
(b) Layer hybridization by intravalley channel is quenched in the neighborhood of Dirac points. Instead, in this low energy sector, layer hybridization is dominated by the \emph{intervalley} Umklapp process.
(c) tBG with $\theta=21.8^{\circ}$, belonging to the chiral point group $D_6$. The red and blue dots represent carbon atoms in the top and bottom layers, respectively. The green and orange shades highlight a subset of vertically aligned carbon atoms, which are linked in a chiral manner by other carbon atoms, reminiscent of the simplified model introduced in section 2.1 (c.f. Fig.~1). 
(d) Band structure of $\theta=21.8^{\circ}$ tBG. Red solid lines are from DFT calculations, and blue dashed lines are the fitting using the atomistic SKTB model (see Appendix B).  
This band structure can be well reproduced by the simplified tight-binding model with the artificial chiral hopping (green dashed lines). 
(e) Topological corner states in the $\theta=21.8^{\circ}$ tBG, with red and blue indicating distribution in the top and bottom layers, respectively. (f) Alignment of the Dirac points at a deviated angle $\theta=21^{\circ}$. Given
the proximity of $\mathbf{K}$ from top and $\mathbf{K}^{\prime}$ from bottom layer on the $\mathrm{C_2}$ circle, \emph{intervalley} Umklapp scattering still dominates in low-energy sector. (g)
Topological corner states from atomistic SKTB calculations at $\theta=21^{\circ}$, essentially unchanged compared to the $\theta=21.8^{\circ}$ case.
}
\label{tBG2D}
\end{figure}

To examine whether such \emph{intervalley} Umklapp interlayer hybridization can capture the structural chiral symmetry and lead to the desired topological chirality, we consider twisted bilayer graphene (tBG) with twist angle $\theta=21.8^{\circ}$ (c.f. Figure~\ref{tBG2D}(c)).
In the absence of interlayer coupling, the Dirac cones at the corners of the BZ from each layer can be folded to either $\mathrm{K_m}$ or $\mathrm{K_m^{\prime}}$ corner of the moir\'{e} Brillouin zone (mBZ) (see Figure~\ref{tBG2D}(a)). 
We analyze the change of electronic structure by the \emph{intervalley} Umklapp interlayer hopping at one of the mBZ corners, comparing with the consequence of the artificial sign-flipped interlayer hopping on the untwisted bilayer structure of $AA$-stacking (c.f. Figure~\ref{TBmodel}b).
We note that the sign-flipped interlayer hopping changes the $AA$-stacked bilayer from a nodal line semimetal to a second-order topological insulator (SOTI), by opening a topological energy gap. 
The SOTI phase is characterized by a nontrivial real Chern number (RCN) $\nu_R$~\cite{YXZhao_2017_PRL,Kruthoff_PRX_2017, bjYang_CPB_2019}, as well as layer-resolved corner states whose real-space chirality is directly controlled by the parameter $\zeta$ (see details in Appendix D). 

We calculate the electronic structure of tBG at $\theta=21.8^{\circ}$, using both density functional theory (DFT) and the atomistic Slater-Koster tight-binding (SKTB) model~\cite{SKTB_2013}.
Results are shown in Figure~\ref{tBG2D}(d). The interlayer coupling by twisting indeed opens a narrow gap of $\sim$2.4 meV near $\mathrm{K}$ point, which is consistent with Ref~\cite{PRL_2008_tBG_Gap}.
Next, we investigate the bulk topological invariant and the bulk-boundary correspondence. 
To study the bulk band topology, we directly compute the RCN $\nu_R$ counting all 56 occupied bands. 
We define $n_{+}^{k_i}$ ($n_{-}^{k_i}$) as the number of occupied bands with positive (negative) $C_{2z}$ eigenvalues at $k_i$. 
Results show that $n_{-}^{\mathrm{M}}=30$ at the $\mathrm{M}$ point and $n_{-}^{\Gamma}=24$ at the $\Gamma$ point, indicating a nontrivial RCN $\nu_R=1$ (see equation(\ref{RCN_Formula}) in the Appendix D). This RCN is consistent with that of the simplified bilayer TB model.
Furthermore, we employ the atomistic SKTB model to demonstrate topological corner states in a large flake of tBG with open boundary condition while maintaining the $C_{6z}$ symmetry. We observe localized corner states (c.f. Figure~\ref{tBG2D}(e)), with layer-resolved chirality determined by the sign of twist angle (see comparison of charge distribution for $\theta=-21.8^{\circ}$ in Figure~\ref{gq} in Appendix F). 
These corner states also fully resemble those in the simplified bilayer TB model with artificial sign-flipped interlayer hopping. Additionally, we find that the parameter $\zeta$ in the simplified bilayer TB model signifies the structural chirality in tBG. Symmetry analysis as well as the correspondence between $\zeta$ and the R- or L-structure are provided in Appendix F.  
Overall, the symmetry, dispersion, and topology of the low-energy physics in tBG at $\theta=21.8^{\circ}$ due to \emph{intervalley} Umklapp interlayer hopping are shown to be equivalent to those of the simplified bilayer model due to the artificial sign-flipped interlayer hopping.

In the following sections~\ref{3DWSM} and~\ref{3DHDS}, the atomistic SKTB calculations are performed  in two types of 3D structures with adjacent layers twisted by $21.8^{\circ}$ (or $-21.8^{\circ}$), to explicitly demonstrate the topological consequences of \emph{intervalley} Umklapp scattering.

\subsection{3D helical graphite as a chiral Weyl semimetal}
\label{3DWSM}

We further substantiate the role of the Umklapp interlayer hopping in 3D twisted structures, as a means to introduce $\pi$-flux $\mathbb{Z}_2$ gauge field and topological chirality.
To realize the 3D chiral Weyl topological semimetal phase in 3D moir\'{e} structures, the required sign-flipped interlayer hopping in the type-I sequence should be achieved in a helical graphite where all adjacent interfaces are twisted by the same angle. 
Namely, one could start with a $AAA$ stacking graphite then rotate each layer by an angle of $n\theta$ around a common hexagon center, where $n$ represents the layer number.
Previous studies have explored the electronic structures of this 3D stacking in the small angle limit~\cite{3DTBG_NL_2019,2019_PRB_MTBG,3DTBG_PRR_2020,TypeB_2024_PRL}.
However, those approaches neglected the \emph{intervalley} Umklapp processes, rendering them inadequate for describing the physics here at $\theta=21.8^{\circ}$.

The 3D helically twisted structure breaks translational symmetry in all spatial directions.
Nevertheless, one can define effective crystal momenta in all three directions, which constitutes a 3D parameter space that resembles the Brillouin zone. First, 
we notice that the system is invariant under a screw rotational operation, i.e. $[\hat{\mathbb{T} }, H]=0$, $\hat{\mathbb{T} }\equiv \hat{R} \hat{T}$, where $\hat{R}$ rotates each layer by $\theta$, and $\hat{T}$ translates it along the out-of-plane $z$ direction by the interlayer distance $d$. 
$\hat{\mathbb{T}}$ has the same algebraic symmetry as the translation operation in periodic structures that underlies the Bloch theorem, which allows us to directly write a generalized Bloch wavefuction 
\begin{equation}
\psi_{k_z}(\mathbf{r})=\frac{1}{\sqrt{N}} \sum_j \text{e}^{\text{i} {k}_{{z}}  (jd)} (\hat{R})^j (\hat{T})^j \phi_0(\mathbf{r}),
\label{generilized_kz}
\end{equation}
where the quantum number $k_z$ represents an effective out-of-plane crystal momentum, and the wavefunction component of the $j$-th layer has been written as $\phi_j = (\hat{R})^j(\hat{T})^j\phi_0(\mathbf{r})$, $\mathbf{r}$ being electron's position vector in-plane.
The layer wavefunction $\phi_0(\mathbf{r})$ is to be solved from the Schr\"{o}dinger equation $\hat{H}_{k_z} \phi_0(\mathbf{r}) = E \phi_0(\mathbf{r})$, where the $k_z$ parameterized Hamiltonian reads,
\begin{equation}
\hat{H}_{k_z} \equiv \left[ \frac{1}{\sqrt{N}} \sum_{j'} \text{e}^{-\text{i} {k}_{{z}}  (j'd)} \hat{R}^{-j'} \hat{T}^{-j'} \right ] \hat{H}  \left[\frac{1}{\sqrt{N}} \sum_j \text{e}^{\text{i} {k}_{{z}}  (jd)} \hat{R}^j \hat{T}^j \right].
\end{equation}

\begin{figure}[t!]
\includegraphics[width=11 cm]{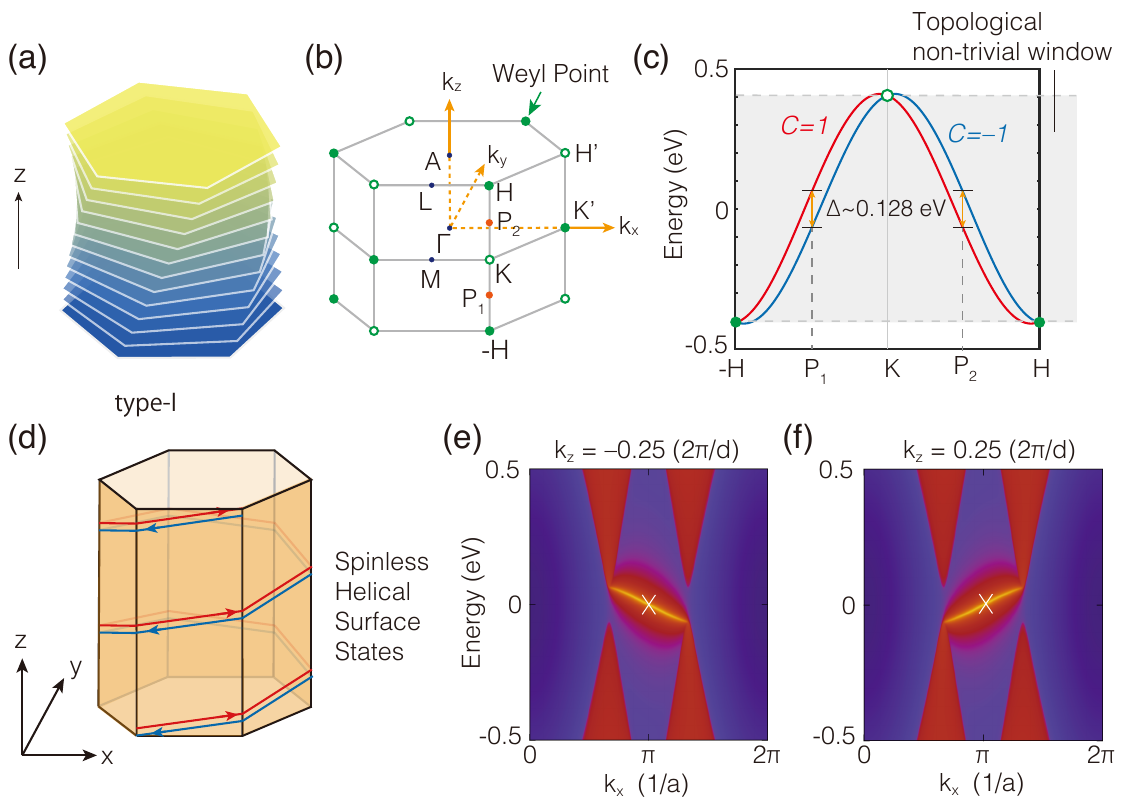}
\centering
\caption{Chiral Weyl semimetal in type-I stacking. 
(a) Illustration of the 3D twisted structure with a constant twist angle $\theta$ between successive layers. 
(b) The corresponding generalized 3D BZ.  
(c) Low-energy band structure. The Chern numbers of each band (red bands: $C=1$, blue bands: $C=-1$).  
(d) Schematic of spinless helical surface states (marked by a white cross in panels (e) and (f)). (e, f) Edge spectra for the 2D subspace with $k_z=\mp 0.25~(2\pi/d)$.}
\label{typeA}
\end{figure}

We note that while each interface has a commensurate interlayer atomic registry, the in-plane periodicity of the interfaces is not aligned, with adjacent ones all having a $21.8^{\circ}$ relative rotation. 
So, this 3D helical stacking does not have in-plane translational symmetry. 
Still, we can introduce effective in-plane crystal momentum $k_x$ and $k_y$ in solving $\phi_0(\mathbf{r})$. 
With interlayer hopping between adjacent layers only, the $k_z$ parameterized Hamiltonian reduces to,
\begin{equation}
\hat{H}_{k_z} =\hat{H} + \textrm{e}^{-\textrm{i} k_z d} \hat{H}\hat{R}^{-1}\hat{T}^{-1} + \textrm{e}^{\textrm{i} k_z d} \hat{H}\hat{R}\hat{T}. 
\label{Hkz}
\end{equation}
With the $21.8^{\circ}$ rotation operator $\hat{R}$ introducing $\sqrt7\times\sqrt7$ moir\'{e} periodicity at each interface, the three terms of ${\hat{H}}_{k_z}$ together have a $7\times7$ periodicity in-plane, from which we can define an effective in-plane crystal momentum $k_x$ and $k_y$ to characterize the layer wavefunction $\phi_0(\mathbf{r})$ (c.f. Appendix E). 
Together with $k_z$, these are the good quantum numbers characterizing the eigenstates of the full Hamiltonian, which span a 3D parameter space of a hexagonal prism shape, i.e. a generalized Brillouin zone as shown in Figure~\ref{typeA}(b). Detailed derivation of the atomistic SKTB model for the twisted structures based on a generalized Bloch theorem is provided in Appendix E. 

By employing the atomistic SKTB method based on a generalized Bloch theorem, we obtain the band structure of the 3D twisted graphite, as shown in Figure~\ref{typeA}(c). 
One observes that the valence and conduction bands touch at  $\mathrm{H}$ and $\mathrm{K}$ points,  which are Weyl nodes with a quantized chiral charge $|C|=1$.
Next, we examine the topological properties of the 2D subsystem $H(k_x, k_y)$ for any fixed value of $k_z$.
For $k_z = 0.25~(2\pi/d)$, a sizeable gap $\sim0.128~\text{eV}$ is observed, which is significantly larger than that in 2D tBG.
Additionally, we observed a topological chiral edge mode in Figure~\ref{typeA}(f), indicating $C=+1$. Further calculations demonstrated that $C=+1$ remains for $k_z > 0$ subsystems, while $C=-1$ for $k_z < 0$ subsystems, as illustrated in Figure~\ref{typeA}(e). 
Note that if we trace the in-gap chiral states marked by white crosses in Figure~\ref{typeA}(e) and (f), topological helical surface states emerge, as shown in Figure~\ref{typeA}(d).
The above demonstration applies to the R-handed 3D helical structure. 
The L-handed structure features reversed Chern numbers, and mirror reflected helical surface states. 
These are the characteristics of a chiral Weyl semimetal~\cite{Chang2018NM,DingHongNC_2019,Hasan2021NRM}, and are consistent with the results from the simplified model with the artificial sign-flipped interlayer hopping (c.f. equation(\ref{typeA-TB})).

\subsection{3D alternating twisted graphite as a higher-order Dirac semimetal}
\label{3DHDS}

\begin{figure}[t!]
\includegraphics[width=11 cm]{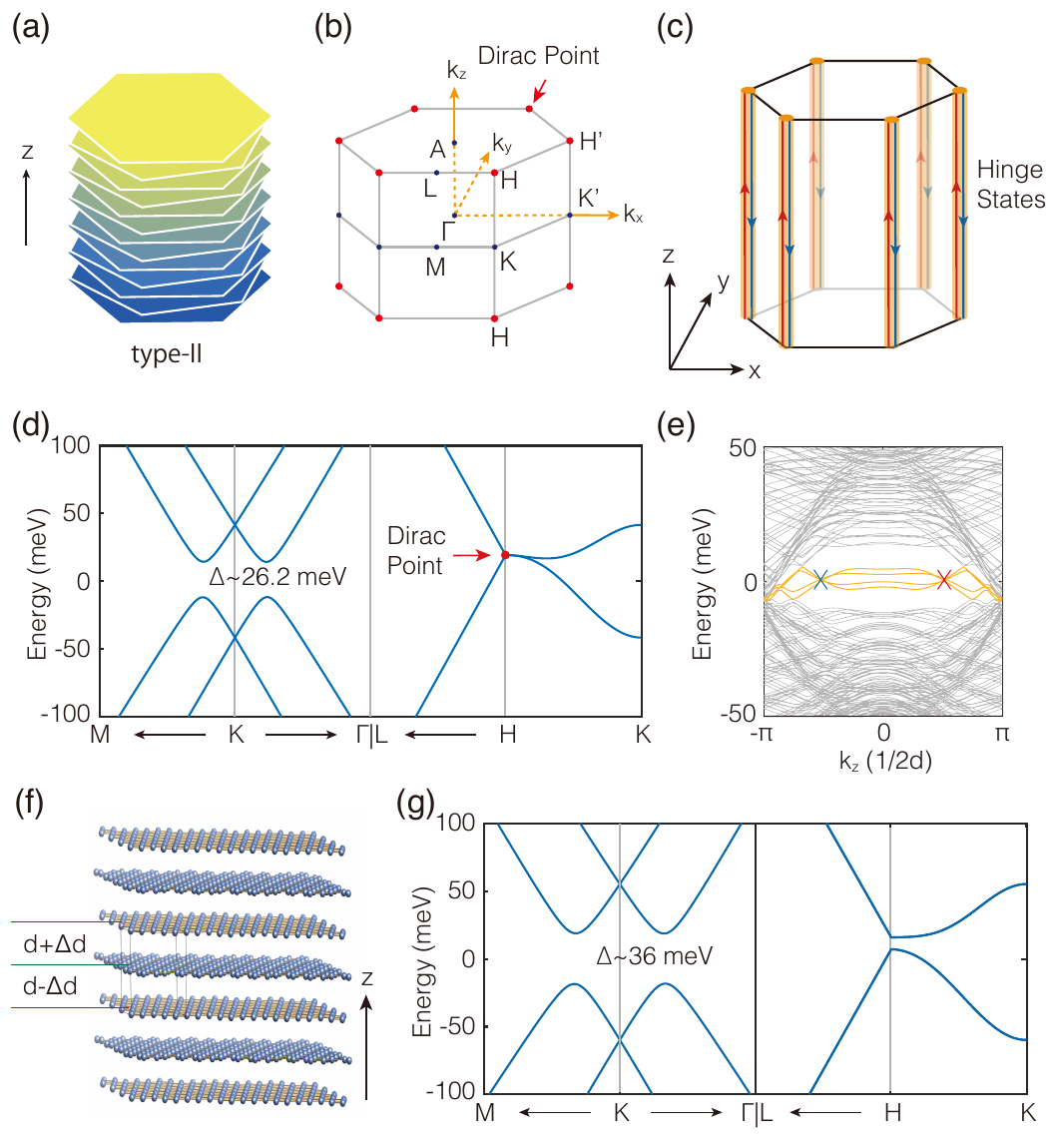}
\centering
\caption{Higher-order topological Dirac semimetal in type-II stacking. 
(a) Illustration of the 3D alternating twisted structure. 
(b) The corresponding BZ. 
(c) Schematic of spinless hinge states (marked by crosses in panel (e)). 
(d) Low-energy band structure, showing the characteristics of a 3D higher-order Dirac semimetal. 
(e) Spectrum for a sample with tubelike geometry as illustrated in (c). The yellow color indicates the in-gap topological hinge bands. The spatial distribution for the states marked by the crosses is shown in (c). 
(f) Schematic of inhomogeneity in the interlayer distances introduced to the structure, which breaks the $M_z$ mirror symmetry.
(g) Low-energy band structure for the type-II stacking with  dimerized interlayer distances:
$d=3.35$~\AA, $\Delta d=0.2$~\AA.
The Dirac point at the H becomes gapped, converting the system into a second-order topological insulator, whereas the topological hinge states remains largely unchanged (c.f. (c)).
}
\label{typeB}
\end{figure}

The type-II model shown in Figure~\ref{TBmodel}(b) can be realized by another type of 3D moir\'{e} structures, i.e. the alternating twisted graphite as shown in Figure~\ref{typeB}(a). In this case, there is bilayer periodicity in $z$ direction, and $\sqrt{7} \times \sqrt{7}$ periodicity in the $x-y$ plane, where the conventional Bloch theorem is applicable.
The crystal structure belongs to the hexagonal space group No. 192. It preserves the same rotational symmetry as graphene, e.g., $C_{2z}$, $C_{6z}$ with respect to $z$-axes. 
Furthermore, spatial inversion symmetry $P$ and time reversal symmetry $T$ are both kept.

The low-energy bulk band structure of 3D type-II twisted graphite is shown in Figure~\ref{typeB}(d), from which one observes a direct band gap $\sim 26.2$ meV near $\mathrm{K}$ (also $\mathrm{K'}$). 
For 2D tBG, the direct band gap is about $\sim 2.4 $ meV, which indicates that interlayer coupling between 2D tBG significantly increases the band gap for 3D tBG. 
Furthermore, one observes a four-fold degenerate Dirac point at $\mathrm{H}$-point. 
Each moir\'{e} BZ contains two Dirac points. 
Remarkably, this is a higher-order topological Dirac semimetal~\cite{wieder2020}, with topological hinge states as shown in Figure~\ref{typeB}(c) and (e), and our calculations are consistent with~\cite{Qian_PRB_2023_Z2}.

The higher-order Dirac semimetal state can be explained by the type-II model, which takes the form
\begin{equation}
      \begin{aligned}
        H_{\text{II}}^{\text{3D}}  = & M \tau_x\sigma_0+M\cos[k_z \cdot (2d)] \tau_x \sigma_0  - M\sin[k_z \cdot (2d)]\tau_y\sigma_0 \\ 
          +& \zeta\lambda(\mathbf{k}_{\|})\left\{ \tau_x \sigma_z + \cos[k_z \cdot (2d)]\tau_x\sigma_z- \sin[k_z \cdot (2d)]\tau_y\sigma_z \right\} \\ 
           +&\chi_1(\mathbf{k}_{\|}) \tau_0 \sigma_x + \chi_2(\mathbf{k}_{\|})\tau_0 \sigma_y,\\
        \end{aligned}
\end{equation}
where $\tau_i$ are the Pauli matrices acting on the layer index. 
Also, we take $\zeta=+$ for simplicity.
When $k_z=0$, $H_{\text{II}}^{\text{3D}}=\chi_1(\mathbf{k}_{\|}) \tau_0  \sigma_x + \chi_2 (\mathbf{k}_{\|}) \tau_0 \sigma_y + 2[M \tau_x \sigma_0 + \lambda(\mathbf{k}) \tau_x \sigma_z]$, representing a reduced 2D bilayer model with enchanced interlayer coupling, which describe a SOTI with a larger band gap and chiral topological corner states.
When $k_z=\pi/(2d)$, $H_{\text{II}}^{\text{3D}}=\chi_1(\mathbf{k}_{\|}) \tau_0  \sigma_x + \chi_2 (\mathbf{k}_{\|}) \tau_0 \sigma_y$, representing a decoupled bilayer graphene system. For $k_z\in ( 0, \pi/(2d))$,  the system retains its 2D SOTI nature with corner states, thereby compromising the topological hinge states.

\subsection{Robustness against symmetry-breaking perturbations}

Topological states in general exhibit robustness against weak symmetry-breaking perturbations, provided that the bulk and edge gaps remain unclosed. From Figure~\ref{typeA}, one observes that the topological edge states emerge in a large gap of $\sim$ 0.128 eV at $k_z=\pm0.25(2\pi/d)$, suggesting robustness of the topological properties. Based on the atomistic SKTB model for 3D twisted structures, we explore below various symmetry-breaking perturbations to examine their impact on the topological state.

{\begin{figure}[t!]
\includegraphics[width=12 cm]{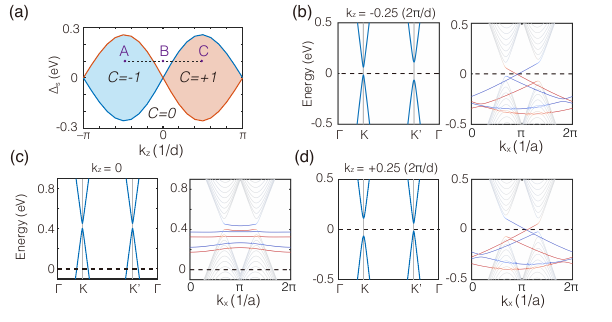}
\centering
\caption{Topological robustness against $C_{2z}$-breaking perturbations. 
(a) Phase diagram of the atomistic SKTB model for 3D helically twisted graphite. (b-d) Bulk and edge spectra for points $\mathrm{A}$, $\mathrm{B}$ and $\mathrm{C}$ in (a). (b) Topological nontrivial with $C=-1$, (c) topological trivial with $C=0$, (d) topological nontrivial with $C=+1$. Note that red and blue lines in edge spectra indicate the edge states from different edges.}
\label{C2zbreak}
\end{figure}

To begin, we consider a $C_{2z}-$breaking staggered potential term $H_s=\left(\Delta_s / 2\right) \sum_{i, \alpha} \xi_i c_{i \alpha}^{\dagger} c_{i \alpha}$, with $\xi_i=\pm1$ for the two sublattices in each layer. For the example of the type-I helically twisted structure, our investigation reveals a phase diagram as a function of this perturbation strength (c.f. Figure~\ref{C2zbreak}), reminiscent of the renowned topological Haldane model~\cite{Haldane_1998}. Here the horizontal axis is $k_z$, playing the role of the artificial magnetic flux in the original Haldane model. From Figure~\ref{C2zbreak}, one finds that, to completely disrupt this topological state, an extraordinarily strong staggered potential exceeding 0.25 eV is needed. 

\begin{figure}[t!]
\includegraphics[width=10 cm]{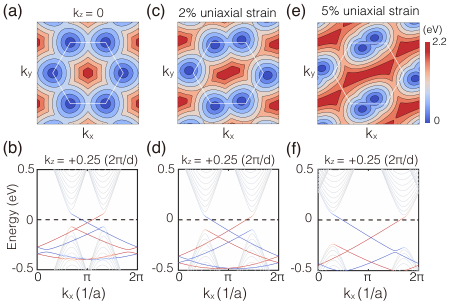}
\centering
\caption{Shift of Weyl point in the type-I twisted structure under uniaxial strain. 
Top row: energy gap contour plots for the conduction and valence bands at each $k$-point in the BZ at $k_z=0$ plane. The white hexagon indicates the first BZ of unstrained structure at $k_z=0$ plane. 
Bottom row: edge spectra of different structures at $k_z=+0.25 (2\pi/d)$. From left to right: structures under 0$\%$, 2$\%$, and 5$\%$ uniaxial strain, respectively.}
\label{C3break}
\end{figure}

Next, we explore the impact of uniaxial strain that breaks the $C_{3z}$ rotational symmetry, also on the example of type-I stacking. 
For each layer, we fix  lattice vector $a_1$ and vary the length of $a_2$, and the screw rotational symmetry is retained. This perturbation does not gap the Weyl points but rather displaces the Weyl points away from the BZ corners on the $k_x-k_y$ plane at $k_z=0$, as shown in the top row of Figure~\ref{C3break}. 
The topological Chern numbers and  edge state's chirality on other gapped $k_x-k_y$ planes at finite $k_z$ remain unaffected. 
Notably, Weyl points persist in pairs and can only be eliminated through pair annihilation. 
Due to their large separation at opposite corners of the mini Brillouin zone, these points are difficult to eliminate by $C_{3z}$-breaking perturbations, even with a uniaxial strain as large as 5$\%$ (see Figure~\ref{C3break}(e) and (f)).

Furthermore, we introduce random in-plane distortions to the type-I twisted structure, where the screw rotational symmetry is still preserved and $\hat{H}_{k_z}$
in equation~(\ref{Hkz}) retains the $7 \times 7$ in-plane periodicity, so that eigenfunctions can still be characterized by the effective crystal momentum, allowing the examination of the robustness of bulk band structures against such distortion. 
All other spatial symmetries are destroyed, as shown in
Figure~\ref{random}(a) where all atoms have been randomly shifted by a magnitude of 0.14~\AA ~in $7 \times 7$ periodic cell.
Remarkably, the Weyl points and topological edge states persist (c.f. Figure~\ref{random}(b-c)). In conclusion, the chiral topological Weyl semimetal phase presents robustness against sizable perturbations due to the significant size of the nontrivial topological band gap and the large separation between Weyl points of opposite chiralities.

\begin{figure}[t!]
\includegraphics[width=10 cm]{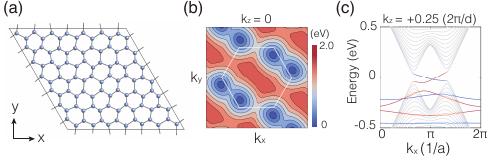}
\centering
\caption{
(a) In-plane distortion where atoms are randomly moved by a magnitude of 0.14~\AA~in a $7 \times 7$ periodic cell (c.f. equation~(\ref{Hkz})). 
(b) Energy gap contour plots for the conduction and valence bands at each $k$-point in the Brillouin zone. The white hexagon indicates the first BZ at $k_z=0$ plane. 
(c) Edge spectra at $k_z=+0.25(2\pi/d)$.}
\label{random}
\end{figure}

The impact of inhomogeneity in the interlayer spacing has also been examined. 
This is computationally overwhelming in type-I stacking, but feasible for atomistic SKTB calculation in the type-II stacking. 
For the 3D alternating twisted graphite, we have considered dimerized interlayer distances (see Figure~\ref{typeB}(f)) with $d=3.35$~\AA, $\Delta d=0.2$~\AA, which breaks $M_z$ mirror symmetry.
We observe that the Dirac point at the H corner becomes gapped (see Figure~\ref{typeB}(g)), which converts the type-II system from a 3D higher-order Dirac semimetal to a second-order topological insulator, whereas the topological hinge states remain largely unchanged.

Lastly, we also find the topological properties to be robust against twist angle deviation from $21.8^{\circ}$. 
Figure~\ref{tBG2D}(f,g) shows the case of a twist angle $\theta=21^{\circ}$. \emph{Intervalley} Umklapp scattering is still the dominating channel for layer hybridization in the low energy sector. 
And our atomistic SKTB calculation for a $21^{\circ}$ twisted bilayer shows the topological corner states are well retained (c.f. Figure~\ref{tBG2D}(g)). 
This points to the robustness of the spinless topological chirality, as such corner states at an individual interface form the basis of the helical surface states in type-I (c.f. Figure~\ref{typeA}(d)) and hinge states (Figure~\ref{typeB}(c)) in type-II 3D  twisted stacking.

\section{Discussion}

In conventional topological phases, the energy window where nontrivial topological properties can be utilized is typically determined by the spin-orbit coupling of limited strength. In contrast, the spectral windows in the spinless topological phases here correspond to interlayer hopping amplitudes. For instance, in chiral topological semimetals, Fermi surfaces with a nontrivial Chern number can form at any energy between the highest and lowest Weyl nodes in a set of connected bands~\cite{Chang2018NM}.
For the spinless chiral Weyl semimetal here, this energy window corresponds to the strength of non-chiral part of interlayer hopping, represented by $M$  in equation(\ref{typeA-TB}), whereas the splitting between bands of opposite Chern number is determined by the strength of chiral part of interlayer hopping, represented by $\lambda_0$ in  equation(\ref{typeA-TB}). The topologically nontrivial spectral window of $\sim$ 0.8 eV is indicated by the dashed gray area in Figure~\ref{typeA}(b).

We note that previous studies have delved into the electronic characteristics of various multilayer or 3D twisted graphite systems~\cite{3DTBG_NL_2019,2019_PRB_MTBG,3DTBG_PRR_2020,TypeB_2024_PRL}. These studies, however, are all in the small angle twist limit though, 
and the key mechanism we explore here, the \emph{intervalley} Umklapp scattering, has been  neglected (and is negligible, as elaborated in appendix C) in  addressing the low energy physics in small angle twist systems. In contrast, in the large-angle twisted regime such as the $21.8^\circ$ case explored here, we have demonstrated the crucial role of the \emph{intervalley} Umklapp scattering in the interlayer hopping that determines the low-energy physics, which underlies the two concrete examples of spinless topological chirality arising solely from structural chirality in three dimensions.

The realization of these spinless topological phases in 3D graphene systems are promised by recent progresses on chemical vapor deposition (CVD) growth of twisted van der Waals structures. An origami-kirigami approach has enabled CVD growth of double-helix structure with arbitrary twist angles that remain uniform across hundreds of graphene layers~\cite{Wang_Spiral_2024,TypeB_2024_PRL}. Away from the screw dislocation, such dual-helical spiral exhibits a 3D periodic structure that just corresponds to the type-II alternating twisted graphite. This may readily allow the exploration of the higher-order Dirac semimetal phase, which has not been observed in realistic electronic materials. 
The growth of continuously twisted super-twisted spirals on non-Euclidean surfaces has also been reported~\cite{Spiral_Science_2020}, shedding light on the realization of the type-I 3D helical graphite. 
Moreover, with twistronics expanding into the realms of photonics and acoustics~\cite{tPhotonics2021APL,tPhotonics2021PRL,tPhotonics2021LSA,twistedTPC_2020,twistedAcoustics2018,tAcoustics2021}, our finding points to new pathways of designing topological chirality from twisted spatial patterning of featureless units (orbitals) in these artificial systems.

\section*{Acknowledgments}
C.C. thanks  Y.X. Zhao and X.-L. Sheng for helpful discussions.
The work is supported by the National Key R\&D Program of China (2020YFA0309600), Research Grant Council of Hong Kong SAR (HKU SRFS2122-7S05, AoE/P701/20, A-HKU705/21), and New Cornerstone Science Foundation.

\appendix
\section*{Appendix A. The first-principles calculations}
\label{App_abinit}
\renewcommand{\theequation}{A.\arabic{equation}}
\setcounter{equation}{0}
\renewcommand{\thefigure}{A\arabic{figure}}
\setcounter{figure}{0}
\renewcommand{\thetable}{A\arabic{table}}
\setcounter{table}{0}

\setcounter{subsection}{0}
\setcounter{figure}{0}
\setcounter{equation}{0}
\setcounter{table}{0}

The first-principles calculations were carried out based on the density-functional theory (DFT), as implemented in the Vienna \textit{ab initio} simulation package (VASP)~\cite{kresse1994,kresse1996}. The ionic potentials were treated by using the projector augmented wave method~\cite{blochl1994}. The band structure results presented in the main text are based on the HSE06 approach~\cite{krukau2006}. The energy cutoff of the plane-wave was set to 500 eV. The energy convergence criterion in the self-consistent calculations was set to \textcolor{black}{10$^{-6}$} eV. A $\Gamma$-centered Monkhort-Pack $k$-point mesh with a resolution of 2$\pi$\texttimes{}0.03\textcolor{black}{{} \AA{}$^{-1}$} was used for the first Brillouin zone sampling.

\section*{Appendix B. The atomistic Slater-Koster tight-binding model of graphite}
\label{App_SKTB}
\renewcommand{\theequation}{B.\arabic{equation}}
\setcounter{equation}{1}
\renewcommand{\thefigure}{B\arabic{figure}}
\setcounter{figure}{1}
\renewcommand{\thetable}{B\arabic{table}}
\setcounter{table}{1}

\setcounter{subsection}{0}
\setcounter{figure}{0}
\setcounter{equation}{0}
\setcounter{table}{0}

The atomistic Slater-Koster tight-binding (SKTB) model~\cite{SKTB_2013} of graphite is given by
\appendix
\begin{equation}
\mathcal{H}=-\sum_{\langle i, j\rangle} t(\mathbf{d}_{i j}) c_i^{\dagger} c_j + h.c.,
\end{equation}
\setcounter{section}{2}
where $c_{i}^{\dagger}$ and $c_{j}$ denote the creation and annihilation operators for the orbital on site $i$ and $j$, respectively, $\mathbf{d}_{ij}$ symbolizes the position vector from site $i$ to $j$, and $t\left(\mathbf{d}_{i j}\right)$ represents the hopping amplitude between sites $i$ and $j$. We adopt the following approximations:
\begin{equation}
\begin{aligned}
-t(\mathbf{d}) & =V_{p p \pi}\left[1-\left(\frac{\mathbf{d} \cdot \mathbf{e}_z}{d}\right)^2\right]+V_{p p \sigma}\left(\frac{\mathbf{d} \cdot \mathbf{e}_z}{d}\right)^2 \\
V_{p p \pi} & =V_{p p \pi}^0 \exp \left(-\frac{d-a_0}{\delta_0}\right) \\
V_{p p \sigma} & =V_{p p \sigma}^0 \exp \left(-\frac{d-d_0}{\delta_0}\right) .
\end{aligned} 
\end{equation}
In the above, $a_0\approx1.42$~\AA~is the nearest-neighbor distance on monolayer graphene, $d_0\approx3.35$~\AA~represents the interlayer spacing, $V_{pp\pi}^0$ is the intralayer hopping energy between nearest-neighbor sites, and $V_{pp\sigma}^0$ corresponds to the energy between vertically stacked atoms on the two layers. 
Here we take $V_{p p \pi}^0\approx-4.32$ eV, $V_{p p \sigma}^0\approx0.78$ eV, 
and $\delta_0=0.45255$~\AA~to fit the dispersions of tBG from DFT result. Hopping for $d>6$~\AA~is  exponentially small and thus neglected in our calculation.

\section*{Appendix C. Difference between the physics of small-angle and large-angle twisted regimes}
\label{App_Umklapp_Diff}
\renewcommand{\theequation}{C.\arabic{equation}}
\setcounter{equation}{2}
\renewcommand{\thefigure}{C\arabic{figure}}
\setcounter{figure}{2}
\renewcommand{\thetable}{C\arabic{table}}
\setcounter{table}{2}

\setcounter{subsection}{0}
\setcounter{figure}{0}
\setcounter{equation}{0}
\setcounter{table}{0}

The interlayer hopping at a twisted interface can be generally described as follows~\cite{tBGHandbook_2019}:
\begin{equation}
T\left(\mathbf{k}_1, \mathbf{k}_2\right)=\frac{1}{S} \sum_{\mathbf{G}_t, \mathbf{G}_{b}} t_{12}\left(\mathbf{k}_{1}+\mathbf{G}_{t}\right) \textrm{e}^{\textrm{i}\left(\mathbf{G}_{t}-\mathbf{G}_{b}\right) \mathbf{R}} \delta_{\mathbf{k}_1+\mathbf{G}_{t}, \mathbf{k}_2+\mathbf{G}_{b}}.
\label{interlayertunneling}
\end{equation}
Electrons can tunnel between two states from top and bottom layer with momentum $\mathbf{k}_1$ and $\mathbf{k}_2$ respectively only if the reciprocal lattice vectors $\mathbf{G}_t$ and $\mathbf{G}_b$ of each layer exist such that
\begin{equation}
\mathbf{k}_1+\mathbf{G}_{t}=\mathbf{k}_2+\mathbf{G}_{b} \Rightarrow \Delta \mathbf{G}=\mathbf{k}_1-\mathbf{k}_2=\mathbf{G}_{b}-\mathbf{G}_{t}=\mathbf{G}_{m},
\label{umklappcondition}
\end{equation}
where $\mathbf{G}_m$ is the reciprocal lattice of the commensuration supercell. 
This is the so-called \emph{generalized Umklapp condition}. Note that Umklapp processes manifest themselves in both small and large angle twisted regimes for tBG~\cite{MacD_PRB_2010,Umk_NP_2019,Umk_Nat_2019,Umk_HongyiPRL2015,Umk_Nat_WY_2019,Umk_2DMater_2020,ishizuka2022wide,Umklapp_SA_2024}, displaying inherent distinctions which will be discussed as follows.
The Umklapp assisted interlayer hopping can be illustrated on the extended zone scheme as shown in Figure~\ref{Umklapp_Appd}(a), where twisting moves the Dirac points from adjacent layers relative to each other on concentric circles. 
We focus on the low energy sector near the Dirac points (denoted by the solid and empty dots), where the effect of interlayer coupling is significant only when the following two conditions are both satisfied: (i) a Dirac point from a layer (blue) is close enough to one from an adjacent layer (red) in the extended zone scheme, so that equation(\ref{umklappcondition}) can be satisfied in the vicinity of the Dirac points; (ii) condition (i) shall be satisfied not too far away from the origin, i.e. on the first several concentric circles, because $t_{12}(\mathbf{k})$ decays fast with $\left|\mathbf{k}\right|$.

\begin{figure}[t!]
\includegraphics[width=12 cm]{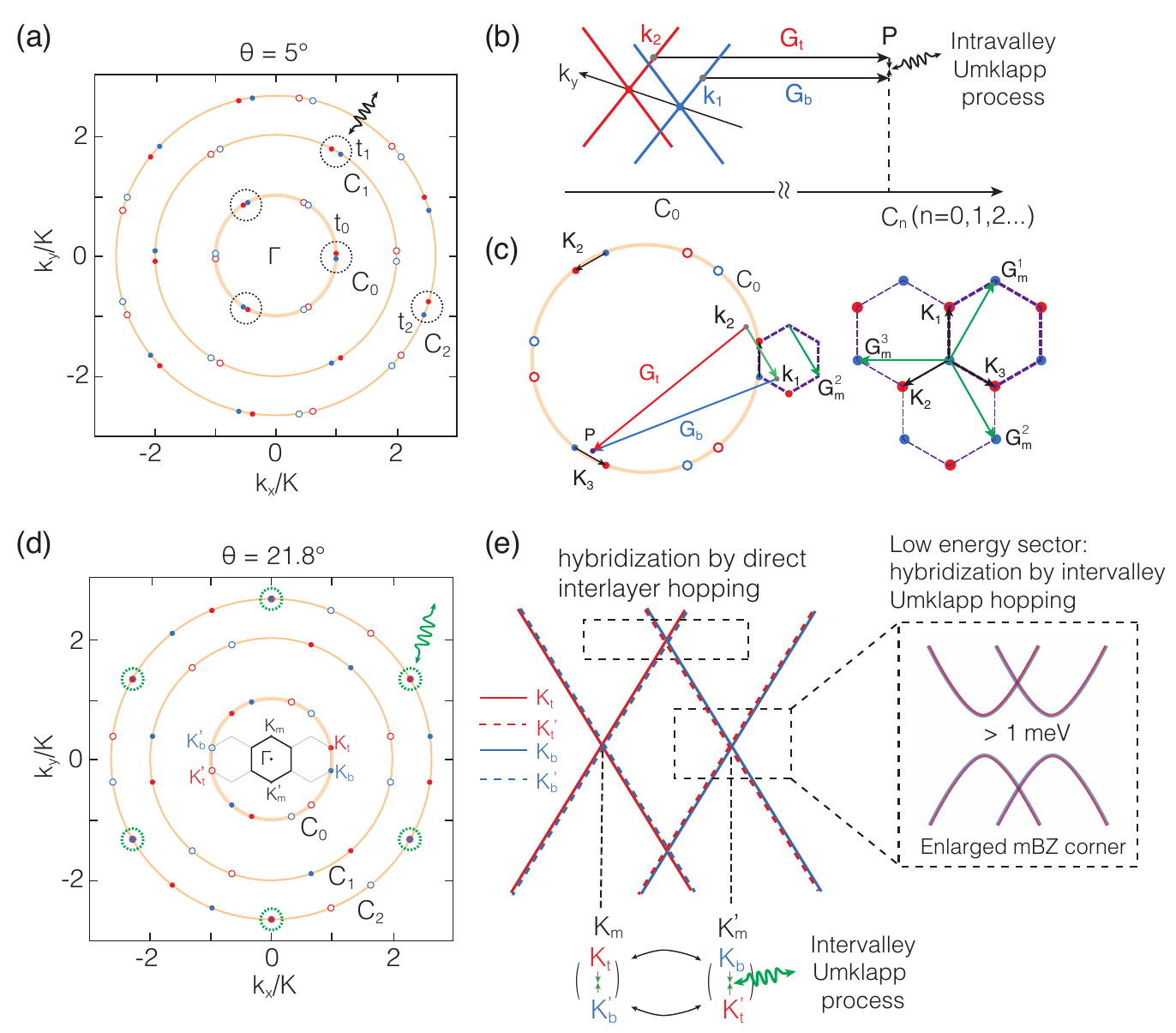}
\centering
\caption{Umklapp assisted interlayer coupling across a twisted interface. 
(a) Schematic figure showing monolayer Dirac points in extended zone scheme, in small twist angle regime. Solid (hollow) dots denote $\mathbf{K}$ ($\mathbf{K}^\prime$) points of the monolayers, and red (blue) color denote top (bottom) layer. 
(b) Schematic figure showing the intravalley Umklapp process. 
(c) Umklapp process in the vicinity of the $\mathrm{C_0}$ circle. 
States from the two layers couple when $\mathbf{k}_1-\mathbf{k}_2=0$, $\mathbf{G}_m^2$, $-\mathbf{G}_m^1$. 
The purple dashed line marks a moir\'{e} unit cell in the reciprocal space. 
Umklapp scattering manifests in a way that a state near the tip of the Dirac cones in one layer are coupled to multiple states of the adjacent layer, aiding layer hybridized electronic states to capture the moir\'{e} pattern periodicity. 
(d) Schematic figure showing monolayer Dirac points for $\theta=21.8^{\circ}$, where $\mathbf{K}$ points of top layer overlap with $\mathbf{K}^\prime$ points of bottom layer on $\mathrm{C_2}$ circle (enclosed by dotted green circles). 
(e) Illustrations depict the interlayer hybridization effect at this large twist angle. Red (blue) color denote top (bottom) layer, while solid (dashed) lines represent bands from their $\mathbf{K}$ ($\mathbf{K}^\prime$) valleys. $\mathrm{K_m}$ and $\mathrm{K_m^{\prime}}$ denote the two moir\'{e} Brillion zone (mBZ) corners. 
Top layer $\mathbf{K}_t$ valley and bottom layer $\mathbf{K}_b^{\prime}$ valley get folded to $\mathrm{K_m}$ mBZ corner where they become degenerate, and \emph{intervalley} Umklapp process opens a hybridization gap at the Dirac point.}
\label{Umklapp_Appd}
\end{figure}

In the small angle twisted regime (Figure~\ref{Umklapp_Appd}(a)), the two conditions are satisfied on circles $\mathrm{C_0}$ and $\mathrm{C_1}$, for the intravalley coupling (i.e. between solid dots, and between empty dots). One can therefore decouple the two valleys, keeping only the leading terms in equation(\ref{interlayertunneling}). This underlies the interlayer coupling for the continuum models in tBG and t-TMDs, which can be written as~\cite{yu2020giant}:
\begin{equation}
T(\mathbf{R})=  t_0 \cdot\left(\textrm{e}^{\textrm{i} \mathbf{K}_1 \mathbf{R}}+\textrm{e}^{\textrm{i} \mathbf{K}_2 \mathbf{R}}+\textrm{e}^{\textrm{i} \mathbf{K}_3 \mathbf{R}}\right)  +t_1 \cdot\left(\textrm{e}^{-2 \textrm{i} \mathbf{K}_1 \mathbf{R}}+\textrm{e}^{-2 \textrm{i} \mathbf{K}_2 \mathbf{R}}+\textrm{e}^{-2 \textrm{i} \mathbf{K}_3 \mathbf{R}}\right)+\text { c.c.}
\label{tunneling}
\end{equation}
Here, with $t_0=t_{12}(G_K)$, $t_1=t_{12}(2G_K)$($G_K=4\pi/3a_0$, $a_0$ is the lattice constant of monolayer), we define $\mathbf{K}_1=\mathbf{K}_t-\mathbf{K}_b$, $\mathbf{K}_2={\hat{C}}_3\mathbf{K}_1$, $\mathbf{K}_3={\hat{C}}_3^{-1}\mathbf{K}_1$ as shown in Figure~\ref{Umklapp_Appd}(c). Considering only $t_0$ terms in equation(\ref{tunneling}) is the first harmonic approximation adopted in many continuum model studies, including that in the Ref.~\cite{3DTBG_PRR_2020}. In this context, only the \emph{intravalley} channels are considered for the interlayer coupling, and Umklapp scattering manifests in a way that a state near the tip of the Dirac cones in one layer are coupled to multiple states of the adjacent layer, \emph{aiding layer hybridized electronic states to capture the moiré pattern periodicity}. For instance, as shown in Figure~\ref{Umklapp_Appd}(c), interlayer hybridization between states $\mathbf{k}_1$ and $\mathbf{k}_2$ can occur under the condition $\mathbf{k}_1-\mathbf{k}_2=\mathbf{G}_t-\mathbf{G}_b=\mathbf{G}_m^2$. This Umklapp process fold the hybridized states from the $\mathbf{P}$ point near $\mathbf{K}_3$ vector to $\mathbf{k}_1$ point within the mBZ.

In the scenario of large twist angles, the Dirac points from the same valley of adjacent layers get widely separated on the first few circles (c.f. Figure~\ref{Umklapp_Appd}(d)). 
So \emph{intravalley} interlayer hopping can now only hybridize states far away from the Dirac points, where Dirac cones from adjacent layers intersect (Figure~\ref{Umklapp_Appd}(e)), and their effects on the low-energy sector near the Dirac points become negligible due to the energy detuning between states that can satisfy the Umklapp condition equation(\ref{umklappcondition}). 
Nonetheless, with the increase of $\theta$, opposite valleys from adjacent layers can approach each other in outer circles, and the larger $\theta$ is, the closer to the origin this can happen. For instance, at commensurate angle $\theta=13.2^{\circ}$, $\mathbf{K}_t$ and $\mathbf{K}_b^{\prime}$ overlap on $\mathrm{C_4}$ circle, while at $\theta=21.8^{\circ}$, $\mathbf{K}_t$ and $\mathbf{K}_b^{\prime}$ overlap on $\mathrm{C_2}$ circle. 
This means that, in the low energy sector, interlayer coupling through the \emph{intervalley} Umklapp channels becomes more and more important with the increase of $\theta$, reaching the maximal effect in the vicinity of $\theta=21.8^{\circ}$. 
It is such \emph{intervalley} Umklapp process that captures the structural chiral symmetry, underlying the topological chirality discussed in the main text.

In Figure~\ref{band_13deg}, we show the calculated electronic structures for twist angle $\theta=13.2^{\circ}$ and $\theta=21.8^{\circ}$. 
One find that, in the low energy sectors, \emph{intervalley} Umklapp coupling opens hybridization gap at the Dirac point: 
at $\theta=13.2^{\circ}$ the hybridization gap is only $\sim$ 0.1 meV, but rapidly increase to $\sim$ 1 meV at $\theta=21.8^{\circ}$. 

\begin{figure}[t!]
\includegraphics[width=10 cm]{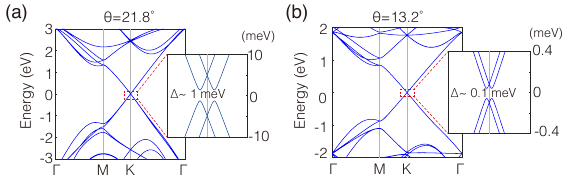}
\centering
\caption{Band structures for tBG at twist angle (a) $\theta=21.8^{\circ}$ and (b) $\theta=13.2^{\circ}$.}
\label{band_13deg}
\end{figure}

\section*{Appendix D. Simplified bilayer model with effective chiral interlayer hopping}
\label{App_AAbilayer}
\renewcommand{\theequation}{D.\arabic{equation}}
\setcounter{equation}{2}
\renewcommand{\thefigure}{D\arabic{figure}}
\setcounter{figure}{2}
\renewcommand{\thetable}{D\arabic{table}}
\setcounter{table}{2}

\setcounter{subsection}{0}
\setcounter{figure}{0}
\setcounter{equation}{0}
\setcounter{table}{0}

For spinless systems with $PT$ symmety, the topology of a 2D insulator is characterized by a $\mathbb{Z}_2$ real Chern number (RCN) $\nu_R$, also known as the second Stiefel-Whitney number~\cite{YXZhao_2017_PRL,Kruthoff_PRX_2017,bjYang_CPB_2019}. 
In 2D systems, when both the $PT$ and $P$ (or $C_{2z}$) symmetries are maintained, calculating the RCN becomes easier and intuitive. One can count the parity eigenvalues of the valence bands at the four inversion-invariant momenta points $\Gamma_{i}$ and apply the formula 
\begin{equation}
(-1)^{v_R}=\prod_{i=1}^4(-1)^{\left\lfloor\left(n_{-}^{\Gamma_i} / 2\right)\right\rfloor},
\label{RCN_Formula}
\end{equation} 
to obtain the RCN $\nu_R$~\cite{YXZhao_2017_PRL,bjYang_CPB_2019}, where $n_{-}^{\Gamma_i}$ represents the number of minus parities in the valence band at $\Gamma_i$.
The presence of a nontrivial RCN $\nu_R=1$ in two copies of graphene suggests that creating a gap in the spectrum of bilayer graphene, such as $AA$-stacked bilayer graphene, holds potential for generating real Chern insulator states.

\begin{figure}[t!]
\includegraphics[width=9 cm]{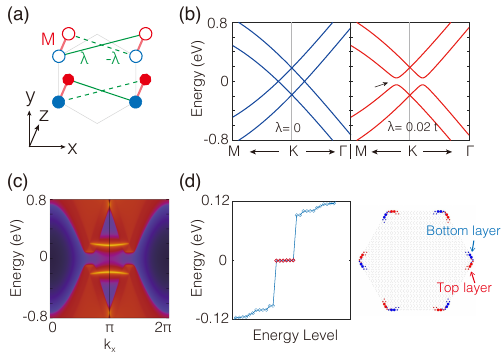}
\centering
\caption{Second-order topological insulator states in the simplified bilayer model with effective chiral interlayer hopping. Solid and dashed lines denote positive and negative hopping amplitudes, respectively. (a) Schematic figure of the $AA$-stacked bilayer graphene and (b) corresponding band structure.  Here, $t_1 = -0.91$ eV, $M = -0.2t_1$, $\lambda_0=0.02t_1$. (c) Projected spectra for zigzag edge. (d) Energy spectrum of the hexagonal-shaped nanodisk shown in the inset. Energy levels are plotted in ascending order. The inset also shows the charge distribution of the states marked in red in the spectra.}
\label{2DTB}
\end{figure}

The simplified bilayer model is therefore constructed by introducing an effective chiral interlayer coupling on top of an $AA$-stacked bilayer graphene lattice. As discussed in the main text, the Hamiltonian in the Bloch basis of $(\psi_{tA}, \psi_{tB},\psi_{bA},\psi_{bB})^{T}$ reads:
\begin{equation}
\begin{aligned}
\mathcal{H}^{2D}_{TB}(\mathbf{k}) & =\chi_1(\mathbf{k}) \tau_0 \sigma_x+\chi_2(\mathbf{k}) \tau_0 \sigma_y+M \tau_x \sigma_0+\zeta\lambda(\mathbf{k}) \text{i} \tau_y \sigma_z, \\
\chi_1 & +\text{i} \chi_2=t_1 \sum_{i=1}^3 \text{e}^{\text{i} \mathbf{k} \cdot \mathbf{\delta}_i}, \\
\lambda(\mathbf{k}) & =2 \text{i} \lambda_0 \sum_{i=1}^3 \sin \left(\mathbf{k} \cdot \mathbf{d}_i\right).
\end{aligned}
\label{2DTB_H}
\end{equation}
Here, $t$ and $b$ denote the layer index, $A$ and $B$ denote the sublattice index, and $\tau_i$ and $\sigma_i$ are the Pauli matrices acting on the layer and sublattice index, respectively. 
The nearest-neighbor intralayer hopping vectors within one layer are given by $\mathbf{\delta}_1=\frac{1}{3} \mathbf{a}_1+\frac{2}{3} \mathbf{a}_2$, $\mathbf{\delta}_2=-\frac{2}{3} \mathbf{a}_1-\frac{1}{3} \mathbf{a}_2$, and $\mathbf{\delta}_3=\frac{1}{3} \mathbf{a}_1-\frac{1}{3} \mathbf{a}_2$. 
The next-nearest interlayer hopping vectors $\mathbf{d}_1=\mathbf{a}_1$, $\mathbf{d}_2=\mathbf{a}_2$, and $\mathbf{d}_3=-\mathbf{a}_1-\mathbf{a}_2$ are also included, with $\zeta=+(-)$. Take $\zeta=+$ for simplicity. 
The Hamiltonian obeys following symmetries $\left\{C_{2 z}, C_{3 z}, T, \mathcal{S}\right\}$($\mathcal{S}= -\tau_z \otimes \sigma_z$ is the sublattice symmetry, which often emerges in carbon allotropes~\cite{Weikang_PRB2018}).
The  sign-flipped interlayer hopping  breaks all the mirror symmetries and spatial inversion symmetry, opening an energy gap in $AA$-stacked bilayer graphene and transforming it to a real Chern insulator.

The band structures with and without the chiral interlayer hopping term are shown in Figure~\ref{2DTB}(b), revealing the gapping of nodal points. 
Remarkably, within the bulk band gap, a pair of gapped edge bands is observed for generic zigzag edges, as depicted in Figure~\ref{2DTB}(c). Next, we investigate the presence of corner states, a key characteristic of a 2D second-order topological insulator (SOTI), we analyze the energy spectrum of a nanodisk as a 0D geometry. 
Specifically, we consider a hexagonal nanodisk, as illustrated in Figure~\ref{2DTB}(d). The resulting discrete energy spectrum, plotted in Figure~\ref{2DTB}(d), reveals the existence of six zero-energy states within the bulk band gap.
Note that these corner states exhibit a distinctive feature compared to those observed in other SOTIs. In this case, the corner states are layer-resolved, manifesting a chiral nature.
Furthermore, this strucrual chirality of corner states can be directly tuned by $\zeta$.

\section*{Appendix E. Atomistic Slater-Koster tight-binding method for the twisted structures based on a Generalized Bloch theory}
\label{App_SKTB_3D}
\renewcommand{\theequation}{E.\arabic{equation}}
\setcounter{equation}{4}
\renewcommand{\thefigure}{E\arabic{figure}}
\setcounter{figure}{0}
\renewcommand{\thetable}{E\arabic{table}}
\setcounter{table}{4}

\setcounter{subsection}{0}
\setcounter{figure}{0}
\setcounter{equation}{0}
\setcounter{table}{0}

\begin{figure}[b!]
\includegraphics[width=10 cm]{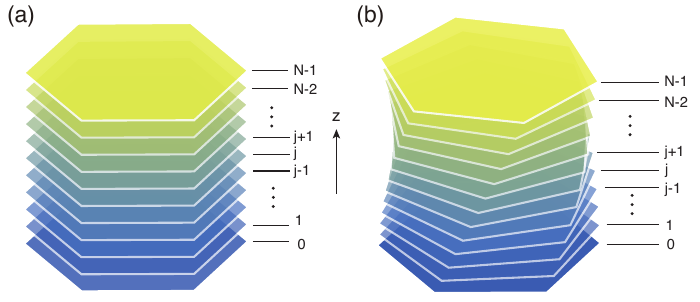}\centering
\caption{ Schematics of two stacking configurations: (a) conventional translational invariant stacking, (b) helical stacking with screw rotational symmetry}.
\label{3D_stacking}
\end{figure}

For an $N$-layered $AA$-stacked system as shown in Figure~\ref{3D_stacking}(a),  the system has a translational symmetry along $z$-direction. 
The wavefunction of the $n$-th layer is  $\phi_n(\mathbf{r})$, where $\mathbf{r}$ is the position vector in-plane. Then the translation operator $T_l$ is defined as $T_l \phi_n = \phi_{n+l}$. $\psi$ is the wavefunction of the system, which is a linear combination of a set of $\phi$. Based on Bloch theorem, it can be known that the eigenvalues of $T_1$ for $\psi$ are $\text{e}^{\text{i} \frac{2\pi}{Nd} md}$, where $m=-\frac{N}{2},\-\frac{N}{2}+1,\ldots, \frac{N}{2}$, $d$ is the interlayer spacing. The Bloch wavefunction is thus given by
\begin{equation}
\psi = \frac{1}{\sqrt{N}}\sum_{n}^{N} \text{e}^{\text{i} \frac{2\pi m n}{N}} \phi_n = \frac{1}{\sqrt{N}}\sum_{n}^{N} \text{e}^{\text{i} n k_m d} \phi_n,
\end{equation}
where $k_m = m \frac{2\pi}{Nd}$.

\begin{figure}[b!]
\includegraphics[width=9 cm]{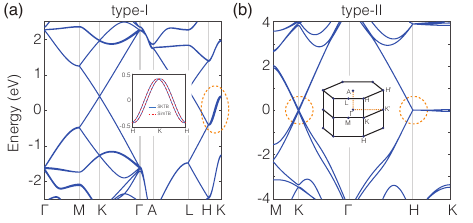}\centering
\caption{Results from atomistic SKTB calculations for 3D twisted graphite. 
Band structures of 3D twisted graphite in (a) type-I stacking and (b) type-II stacking. 
The yellow dashed circles indicate the specific regions of focus discussed in the main text. 
The inset in (a) shows the fitting of the simplified TB model with chiral interlayer hopping (red dashed lines). 
The coordinates of high symmetry points in 3D BZ are given by $\Gamma(0,0,0)$,$\mathrm{M}(0,1/2,0)$, $\mathrm{K}(1/3,1/3,0)$,$\mathrm{A}(0,0,1/2)$, $\mathrm{L}(0,1/2,1/2)$, and $\mathrm{H}(1/3,1/3,1/2)$ in reciprocal lattice.}
\label{typeAB_bulk}
\end{figure}

For an $N$-layered helical stacking system as shown in Figure~\ref{3D_stacking}(b), the structure exhibits a screw rotational symmetry, denoted as $\mathbb{T}_l$, which consists of an in-plane $l\theta$ rotation followed by an out-of-plane translation of $ld$. We have $\mathbb{T}_l\phi_j=\phi_{j+l}$ and $\left[ \mathbb{T}_l, H\right]=0$. The wavefunction of the $j$-th layer is $\phi_j=(\hat{R})^j (\hat{T})^j \phi_0$, where $\hat{R}$ ($\hat{T}$) is a rotation (translational) operation, and its superscript indicates how many times the operation has been performed. 
One notice that a group of $\left\{T_l\right\}$ is isomorphism to a group of $\left\{ \mathbb{T}_l\right\}$.
Therefore, the eigenstates of $\mathbb{T}_1$ for $\psi$ can be directly obtained by $\text{e}^{\text{i} \frac{2\pi}{Nd} md}$.  Thus the generalized Bloch wavefuction of the system is given by
\begin{equation}
\psi_{k_z}(\mathbf{r})=\frac{1}{\sqrt{N}} \sum_j \mathrm{e}^{\mathrm{i} k_z(j d)}(\hat{R})^j(\hat{T})^j \phi_0(\mathbf{r}),
\end{equation}
where the quantum number $k_z$ represents an effective out-of-plane crystal momentum, and the wavefunction component of the $j$-th layer has been written as $\phi_j = (\hat{R})^j(\hat{T})^j\phi_0(\mathbf{r})$.
The layer wavefunction $\phi_0(r)$ is to be solved from the Schrodinger equation $\hat{H}_{k_z} \phi_0(\mathbf{r}) = E \phi_0(\mathbf{r})$, where the $k_z$ parameterized Hamiltonian reads,
\begin{equation}
\hat{H}_{k_z} \equiv \left[ \frac{1}{\sqrt{N}} \sum_{j'} \text{e}^{-\text{i} {k}_{{z}}  (j'd)} \hat{R}^{-j'} \hat{T}^{-j'} \right ] \hat{H}  \left[\frac{1}{\sqrt{N}} \sum_j \text{e}^{\text{i} {k}_{{z}}  (jd)} \hat{R}^j \hat{T}^j \right].
\end{equation}
We note that while each interface has a commensurate interlayer atomic registry, the in-plane periodicity of the interfaces is not aligned, with adjacent ones all having a $21.8^{\circ}$ relative rotation. 
So, this 3D helical stacking does not have in-plane translational symmetry. 
Still, we can introduce effective in-plane crystal momentum $k_x$ and $k_y$ in solving $\phi_0(\mathbf{r})$. 
With interlayer hopping between adjacent layers only, the $k_z$ parameterized Hamiltonian reduces to,
\begin{equation}
\hat{H}_{k_z} =\hat{H} + \textrm{e}^{-\textrm{i} k_z d} \hat{H}\hat{R}^{-1}\hat{T}^{-1} + \textrm{e}^{\textrm{i} k_z d} \hat{H}\hat{R}\hat{T}. 
\label{Hkz_App}
\end{equation}
With $\sqrt7\times\sqrt7$ moir\'{e} periodicity at each interface, ${\hat{H}}_{k_z}$ therefore has a $7\times7$ periodicity in-plane, from which we can define an effective in-plane crystal momentum $\mathbf{k}_{\|}=(k_x, k_y)$. Therefore, we define the Bloch function by considering a three-layer moir\'{e} periodicity, given by
\begin{equation}
 \phi_0(\mathbf{k_{\|}}) = \frac{1}{\sqrt{N_1/X}} \frac{1}{\sqrt{N_2/X}} \sum_{\mathbf{R}^{S}_{l}} \text{e}^{\text{i} \mathbf{k_{\|}} \mathbf{R}^{S}_l } D_{m,\mathbf{R}_{i,j}},
\end{equation}
where $m$ represents the $A$ and $B$ sublattices, $\mathbf{R}_{i,j}$ denotes the indices of the original graphene unit cell, $\mathbf{R}_{i,j} = i \mathbf{a}_1 + j \mathbf{a}_2$ with $i,j = 0,1,2\cdots X-1$ ($X=7$ for $\theta=21.8^{\circ}$). $\mathbf{R}^{S}_l$ represents the supercell defined by the three-layer moir\'{e} lattice. 
$D_{m,\mathbf{R}_{i,j}} = D_{m}(\mathbf{r}_m-\mathbf{R}_{i,j})$ represents the Wannier function of the $m$ sublattice  at the $\mathbf{R}_{i,j}$ unit cell. 
Similarly, we have a parameterized Hamiltonian of the form
\begin{equation}
\hat{H}_{k_z, \mathbf{k}_{\|}}=  \frac{X^2}{N_1 N_2} \sum_{l, l^{\prime}}\left[ \textrm{e}^{-\textrm{i} \mathbf{k}_{\|} \mathbf{R}_{l^{\prime}}^{S}} \hat{H}_{k_z} \textrm{e}^{\textrm{i} \mathbf{k}_{\|} \mathbf{R}_{l}^{S}} + \mathrm{e}^{\mathrm{i} \mathbf{k}_{\|}\left(\mathbf{R}_l^S-\mathbf{R}_{l^{\prime}}^S\right)} \textrm{e}^{-\textrm{i} k_z d} \hat{H}\hat{R}^{-1}\hat{T}^{-1} + \mathrm{e}^{\mathrm{i} \mathbf{k}_{\|}\left(\mathbf{R}_l^S-\mathbf{R}_{l^{\prime}}^S\right)} \textrm{e}^{\textrm{i} k_z d} \hat{H}\hat{R}\hat{T}\right].
\end{equation}
The full band structures of 3D graphite in type-I and type-II stacking from the atomistic SKTB calculations are shown in Figure~\ref{typeAB_bulk}.

\section*{Appendix F. Comparison of results from atomistic SKTB model and the simplified TB model with effective chiral interlayer hopping}
\label{App_Comparison}
\renewcommand{\theequation}{F.\arabic{equation}}
\setcounter{equation}{3}
\renewcommand{\thefigure}{F\arabic{figure}}
\setcounter{figure}{3}
\renewcommand{\thetable}{F\arabic{table}}
\setcounter{table}{3}

\setcounter{subsection}{0}
\setcounter{figure}{0}
\setcounter{equation}{0}
\setcounter{table}{0}

\begin{figure}[t!]
\includegraphics[width=12 cm]{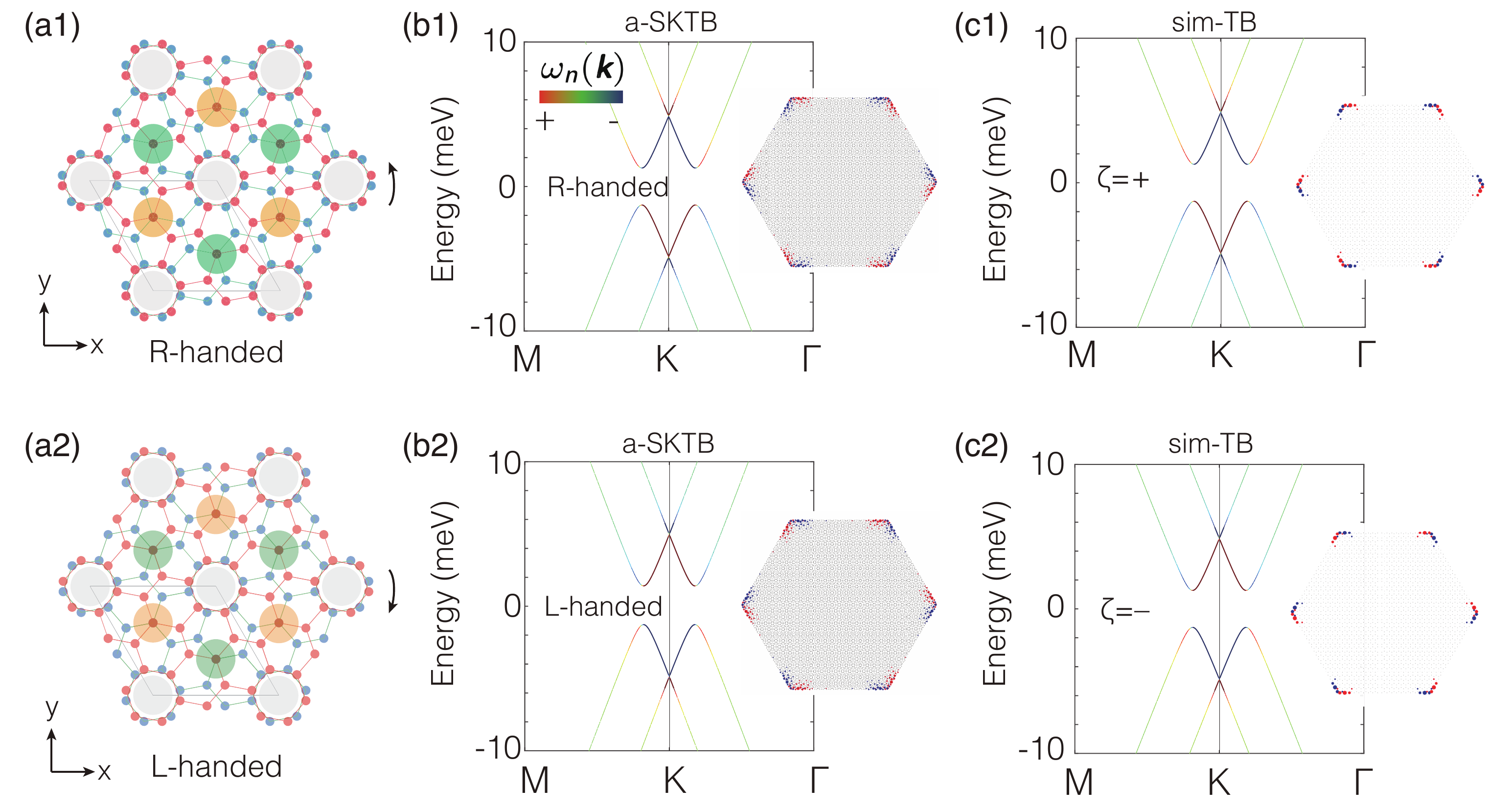}
\centering
\caption{ Comparison of results from atomistic SKTB model and the simplified bilayer TB model with chiral interlayer hopping. The lattice structure of tBG with a commensurate twist angle of $\theta=21.8^{\circ}$ is shown in (a1) and (a2), corresponding to R-handed and L-handed configurations, respectively. Low-energy band strucures from the a-SKTB model (b) and the simplifield bilayer TB model (c), color coding denotes $\mathbf{k}$-space vorticity $\omega_n(\mathbf{k})$. The inset shows the charge distribution of topological corner states.  The top row and bottom row are mirror images of each other. }
\label{gq}
\end{figure}

\begin{figure}[t!]
\includegraphics[width=12 cm]{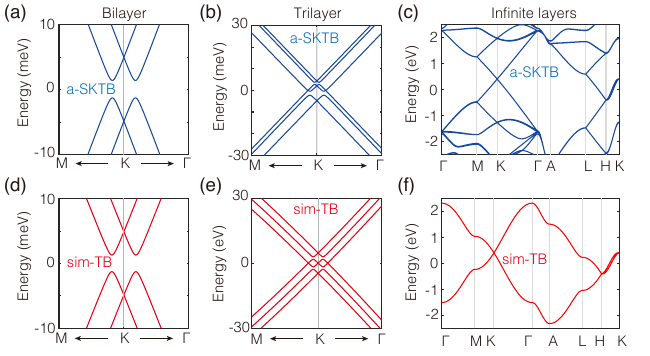}
\centering
\caption{ Comparison of band structures using the simplified TB (sim-TB) model with effective chiral interlayer hopping on an untwisted hexagonal lattice (see section 2.1) and the atomistic SKTB model for twisted stacking. The top row displays band structures for the atomistic SKTB model from bilayer (a) to trilayer (b) to infinite layers (c). The bottom row illustrates band structures for the sim-TB model from bilayer (d) to trilayer (e) to infinite layered systems (f).}
\label{2Dto3D}
\end{figure}

The simplified tight-binding model with effective chiral interlayer hopping utilized in sections 2.1 and 2.2 effectively captures the low-energy physics of both bilayer and 3D twisted graphite. In the following, we delve into a comparative analysis of these two models in bilayer, trilayer, and multilayer systems.

First, consider bilayer system, we find that the parameter $\zeta$ in the bilayer model represents the structural chirality in tBG. The band structures of the $AA$-stacked bilayer model are the same for $\zeta=+(-)$, and the band structures of the two enantiomers in tBG are also identical. This naturally suggests a connection between the structural chirality and the parameters $\zeta$, which we will establish as follows. 
Firstly, we note that this relationship also holds for 2D systems: $M_y \mathcal{H}^{2D}_{TB}(\zeta) M_y^{-1}=\mathcal{H}^{2D}_{TB}(-\zeta)$. 
This implies that reversing the sign of $\zeta$ is equivalent to a spatial mirror reflection. 
Then, we can establish a clear correspondence between $\zeta$ and the R- or L-structure. 
To do so, we conduct a comprehensive comparison of the band geometric quantity and the distribution of corner states obtained from the $AA$-stacked bilayer model and the Slater-Koster tight-binding (SKTB) method~\cite{SKTB_2013} for different handednesses. The comparison of energy bands and distribution of topological corner states from the SKTB model and $AA$-stacked bilayer model is depicted in Figure~\ref{gq}.
The color coding denotes $k$-space vorticity $\omega_n(\mathbf{k})$, which serves as a band geometric quantity of layer current, as expressed in the form~\cite{NC_2023_CXiao}
\begin{equation}
\omega_n(\mathbf{k})=\hbar \operatorname{Re} \sum_{n_1 \neq n} \frac{\left[\mathbf{v}_{n n_1}(\mathbf{k}) \times \mathbf{v}_{n_1 n}^{\text {sys }}(\mathbf{k})\right]_z}{\varepsilon_n(\mathbf{k})-\varepsilon_{n_1}(\mathbf{k})},
\end{equation}
where $n$ and $\mathbf{k}$ represent the band index and crystal momentum, respectively. The term $\mathbf{v}^{\text{sys}}_{n_1 n}(\mathbf{k})=\left\langle u_{n_1}(\mathbf{k})\left|\frac{1}{2}\left\{\hat{\mathbf{v}}, \hat{P}^{\text {sys}}\right\}\right| u_{n}(\mathbf{k})\right\rangle$ involves the operator $\hat{P}^{\text {sys }}=\left(1+\hat{l}^z\right) / 2$, with $\hat{l}^z=\operatorname{diag}(1,-1)$. This operator helps to distinguish between the two enantiomers as it carries information about the layer degree.
The results obtained from both methods are consistent, as shown in Figure~\ref{gq}. Additionally, it can be observed that a positive value of $\zeta$ in the $AA$-stacked bilayer model corresponds to a R-handed structure, whereas a negative value of $\zeta$ corresponds to a L-handed structure. 

Figure~\ref{2Dto3D} further illustrates the comparison of the simplified TB model and atomistic SKTB calculations of type-I twisted structures, for the cases of trilayer and 3D bulk with infinite numbers of layers. The sim-TB model can indeed capture the low-energy physics of twisted structures from few layer limit to 3D bulk.

\section*{Appendix G. From surface states to corner states at few layer limit}
\label{App_Comparison}
\renewcommand{\theequation}{G.\arabic{equation}}
\setcounter{equation}{3}
\renewcommand{\thefigure}{G\arabic{figure}}
\setcounter{figure}{3}
\renewcommand{\thetable}{G\arabic{table}}
\setcounter{table}{3}

\setcounter{subsection}{0}
\setcounter{figure}{0}
\setcounter{equation}{0}
\setcounter{table}{0}

\begin{figure}[t!]
\includegraphics[width=12 cm]{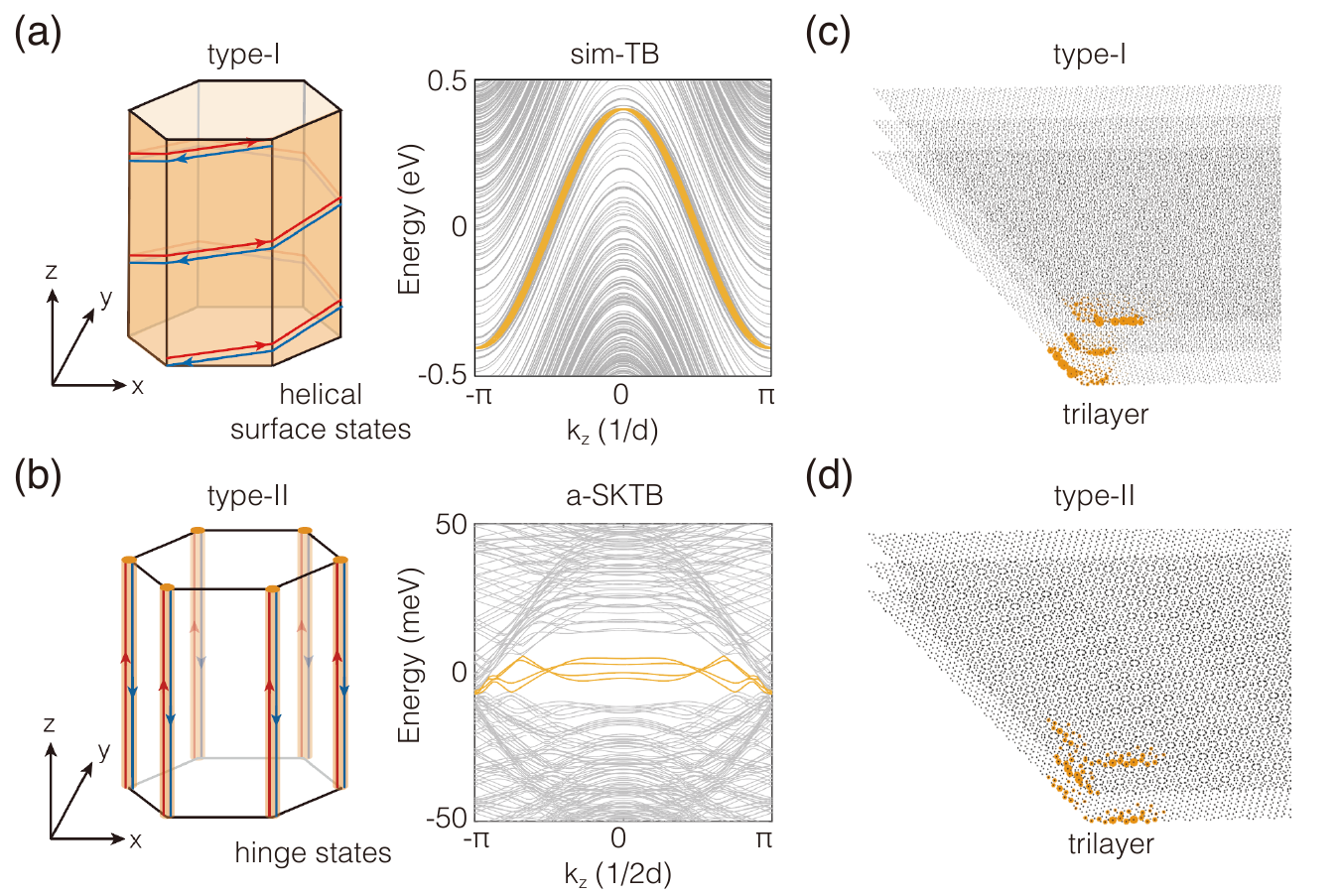}
\centering
\caption{ (a, b) Energy spectra with opend boundary condition in $x-y$ plane and periodic boundary condition in $z$ direction for (a) type-I and (b) type-II stacking. The yellow color indicates the spatial profile of helical/hinge topological surface states. (c) and (d) show the topological corner states for trilayer twisted structures of the two types. The spectrum in part (a) has used the simplified tight-binding model (c.f. Section 2.1). (b,c,d) are from the atomistic SKTB calculations of the twisted structures.}
\label{sizeeffect}
\end{figure}

At finite thickness. The topological helical type surface states in type-I and hinge type surface states in type-II structure persist, subject to the finite size effect in the $z$ direction. In the infinite layer limit, the helical/hinge surface states characterized by the $k_z$ quantum number are running waves (c.f. Figure~\ref{sizeeffect}(a,b)). With finite number of layers, the finite size effect in $z$ direction will hybridize waves running in the $+z$ and $-z$ directions into standing waves, and both the bulk states and the surface states can become gapped at small enough thickness. The ultimate thin limit of both twisted structures is the twisted bilayer, where the surface state manifests as the corner states in this extreme finite size limit (c.f. Figure~\ref{tBG2D}(c-e)). In Figure~\ref{sizeeffect}, we show corner states of trilayer structures, where difference between the type-I and type-II twisted stacking develop. And increasing the number of layers will allow experimental demonstration of how these corner states gradually develop into helical and hinge type respectively in the two types of twisted stacking.

\bibliographystyle{iopart-num}


\end{document}